\documentclass[fleqn,useAMS,usenatbib]{mnras}

\usepackage[T1]{fontenc}

\DeclareRobustCommand{\VAN}[3]{#2}
\let\VANthebibliography\thebibliography
\def\thebibliography{\DeclareRobustCommand{\VAN}[3]{##3}\VANthebibliography}

\usepackage{graphicx}	% Including figure files
\usepackage{amsmath}	% Advanced maths commands
\usepackage{xspace}     % Space after a command
\usepackage{amssymb}    % Required for varpi
\usepackage{ulem}

\newcommand{\gaia}{{\it Gaia}\xspace}
\newcommand{\kms}{{$\text{km}\,\text{s}^{-1}$}\xspace}

\newcommand{\bprp}{{${\rm G_{BP}-G_{RP}}$}\xspace}

\newcommand{\sssss}{{S$^5$}\xspace}
\newcommand{\INT}{{\it INT}\xspace}
\newcommand{\NTT}{{\it NTT}\xspace}

\title[Improved constraints on GC ejection of HVSs]{Improved constraints on Galactic Centre ejection of hypervelocity stars based on novel search method}

\author[S. Verberne et al.]{
Sill Verberne,$^{1}$\thanks{E-mail: verberne@strw.leidenuniv.nl}
Elena Maria Rossi,$^{1}$
Sergey E.~Koposov,$^{2,3,4}$
Tommaso Marchetti,$^{5}$
Konrad Kuijken,$^{1}$
\newauthor
Zephyr Penoyre,$^{1}$
Fraser A.~Evans,$^{6,7}$
Dimitris Souropanis,$^{8,9}$
and Cl\'{a}r-Br\'{i}d Tohill,$^{8,10}$\\
$^{1}$Leiden Observatory, Leiden University, P.O. Box 9513, 2300 RA Leiden, the Netherlands\\
$^{2}$Institute for Astronomy, University of Edinburgh, Royal Observatory, Blackford Hill, Edinburgh EH9 3HJ, UK\\
$^{3}$Institute of Astronomy, University of Cambridge, Madingley Road, Cambridge CB3 0HA, UK\\
$^{4}$Kavli Institute for Cosmology, University of Cambridge, Madingley Road, Cambridge CB3 0HA, UK\\
$^{5}$European Southern Observatory, Karl-Schwarzschild-Strasse 2, 85748 Garching bei Munchen, Germany\\
$^{6}$David A. Dunlap Department of Astronomy and Astrophysics, University of Toronto, 50 St. George Street, Toronto, ON M5S 3H4, Canada\\
$^{7}$Dunlap Institute for Astronomy and Astrophysics, University of Toronto, 50 St. George Street, Toronto, ON M5S 3H4, Canada\\
$^{8}$Isaac Newton Group of Telescopes, Apartado 321, E-38700 Santa Cruz de La Palma, Canary Islands, Spain\\
$^{9}$Institute of Astrophysics FORTH, 71110 Heraklion, Greece\\
$^{10}$University of Nottingham, School of Physics \& Astronomy, Nottingham, NG7 2RD, UK
}

\date{Accepted XXX. Received YYY; in original form ZZZ}

\pubyear{2024}

\begin{document}
\label{firstpage}
\pagerange{\pageref{firstpage}--\pageref{lastpage}}
\maketitle

\begin{abstract}
% It should be a single paragraph not more than 250 words (200 words for Letters).
Hypervelocity stars (HVSs) are stars which have been ejected from the Galactic Centre (GC) at velocities of up to a few thousand \kms. They are tracers of the Galactic potential and can be used to infer properties of the GC, such as the initial-mass function and assembly history. HVSs are rare, however, with only about a dozen promising candidates discovered so far. In this work we make use of a novel, highly efficient method to identify new HVS candidates in \gaia. This method uses the nearly radial trajectories of HVSs to infer their distances and velocities based on their position and \gaia proper motion alone. Through comparison of inferred distances with \gaia parallaxes and photometry we identified 600 HVS candidates with G<20 including the previously discovered S5-HVS1, out of which we obtained ground-based follow-up observations for 196 stars. As we found no new HVSs based on their radial velocity, we used detailed HVS ejection simulations to significantly improve previous HVS ejection rate constraints. In particular, the ejection rate of HVSs more massive than 1 $\mathrm{M_\odot}$ cannot be higher than $10^{-5}$ yr$^{-1}$ at $2\sigma$ significance. Additionally, we predict that there are 5--45 unbound HVSs in the complete \gaia catalogue ($1\sigma$ interval), most of which will be main-sequence stars of a few M$_\odot$ at heliocentric distances of tens to hundreds of kpc. By comparing our results to literature HVS candidates, we find an indication of either a time-dependent ejection rate of HVSs or a non-GC origin of previously identified HVS candidates. 

\end{abstract}


%%%%%%%%%%%%%%%%%%%%%%%%%%%%%%%%%%%%%%%%%%%%%%%%%%

%%%%%%%%%%%%%%%%% BODY OF PAPER %%%%%%%%%%%%%%%%%%

\section{Introduction}
The Galactic Centre (GC) is a highly complex environment, hosting both a Nuclear Star Cluster and a super-massive black hole called \textit{Sagittarius A$^*$} \citep[Sgr A$^*$; e.g.][]{Genzel_2010}. The combination of these two factors gives rise to a unique, high-energy environment within our Galaxy, which challenges our understanding of key physical processes. The origin of the S-star cluster, for instance, remains unknown to this day, because the strong tidal force in this region inhibits standard star formation from molecular clouds \citep{Genzel_2010}. In addition, the initial-mass function (IMF) of stars near the GC has been a long-standing point of debate, whose resolution can give us an important clue as to the mass assembly hisotry of the GC. Some studies find it is consistent with the canonical IMF from \citet{Kroupa_2001} \citep{Maness_2007, Lockmann_2010}, while others find evidence for a top-heavy IMF in disc structures near Sgr A$^*$ \citep[e.g.][]{Paumard_2006, Klessen_2007, Bartko_2010, Lu_2013, Fellenberg_2022}. The combination of high line-of-sight extinction and source crowding make the study of the GC challenging, requiring specialised instruments such as \textit{GRAVITY} \citep{Eisenhauer_2011}.

One promising means of better understanding our GC comes from hypervelocity stars (HVSs), which are stars ejected from the GC at extremely high velocities. They are believed to be ejected from the vicinity of Sgr A$^*$ (order of $10$ AU), but can be observed in parts of the sky more accessible to study \citep{Hills_1988, Yu_2003}. This in turn allows for detailed stellar parameter measurements with existing observatories, and the study of wavelength ranges entirely inaccessible near the GC. It has also been suggested that HVSs could explain the existence of the S-star cluster, because the companion of the HVS in the progenitor binary is deposited at radii consistent with the S-star cluster \citep{Gould_2003}. This migration of already-formed stars into the S-star cluster could therefore solve the issue of in-situ star formation in the strong tidal field near Sgr A$^*$ \citep[e.g.][see, however, \citet{habibi_17}]{Ghez_2003}.

In recent years, the term HVS has been used inconsistently. For clarity, we define an HVS to be characterised exclusively by having been ejected from the vicinity of Sgr A$^*$. Since ejections can occur at a range of velocities, some of these HVSs will still be bound to the Galaxy.  

The discovery of HVSs has proved challenging in spite of their significant scientific potential. Only a few dozen promising candidates \citep[e.g.][]{Brown_2014} and a single star that can be confidently traced back to the GC \citep{Koposov_2020} have been identified. The main challenges include their rarity and the difficulty in disentangling these stars from e.g.~hyper runaway stars from the Galactic disc \citep[e.g.][]{Przybilla_2008, Kreuzer_2020, Irrgang_2021}. 

In this work we perform a targeted survey of HVS candidates. The candidates are identified using a novel, highly efficient selection method with \gaia Data Release 3 \citep[DR3;][]{Gaia_2016, Gaia_2023}, which means we have excellent sensitivity to HVSs from relatively few candidates. We perform ground-based follow-up observations for the most promising HVS candidates and use these observations in combination with sophisticated models to constrain the GC environment and the ejection of HVSs. We use these constraints to investigate the total population of HVSs accessible to \gaia in current and future data releases. We furthermore evaluate previously identified candidates using our observational constraints.

In Section~\ref{sec:selection} we describe how we select HVS candidates from \gaia DR3. In Section~\ref{sec:observations} we present our follow-up observations. In Section~\ref{sec:model} we discuss the model and simulations, in addition to how we apply the observational selections to the models. In Section~\ref{sec:constraints} we present constraints on the GC environment and HVS from our observations in combination with the simulations. In Section~\ref{sec:discussion} we discuss our results and implications for existing HVS candidates as well as what we might expect from \gaia DR4. Finally, in Section~\ref{sec:conclusion} we provide closing remarks.

\section{HVS candidate selection}
\label{sec:selection}
The main challenge when searching for HVSs is that they are extremely rare; there are for instance no HVSs with velocity greater than 700 km s$^{-1}$ in the 34M sources in \gaia DR3 with radial velocities, that could be robustly identified \citep{Marchetti_2022}. HVS candidates selected from observational catalogues for follow-up observations are therefore typically strongly contaminated by non-HVSs. Here we present our new method to identify HVS candidates originating from the GC. In Section~\ref{sec:radial_trajectory} we present the general principle we use to find HVSs, and in Section~\ref{sec:selection_function} we provide the precise selections on the \gaia catalogue for our observational campaign.

\subsection{Radial trajectory}
\label{sec:radial_trajectory}

\begin{figure}
    \centering
    \includegraphics[width=\linewidth]{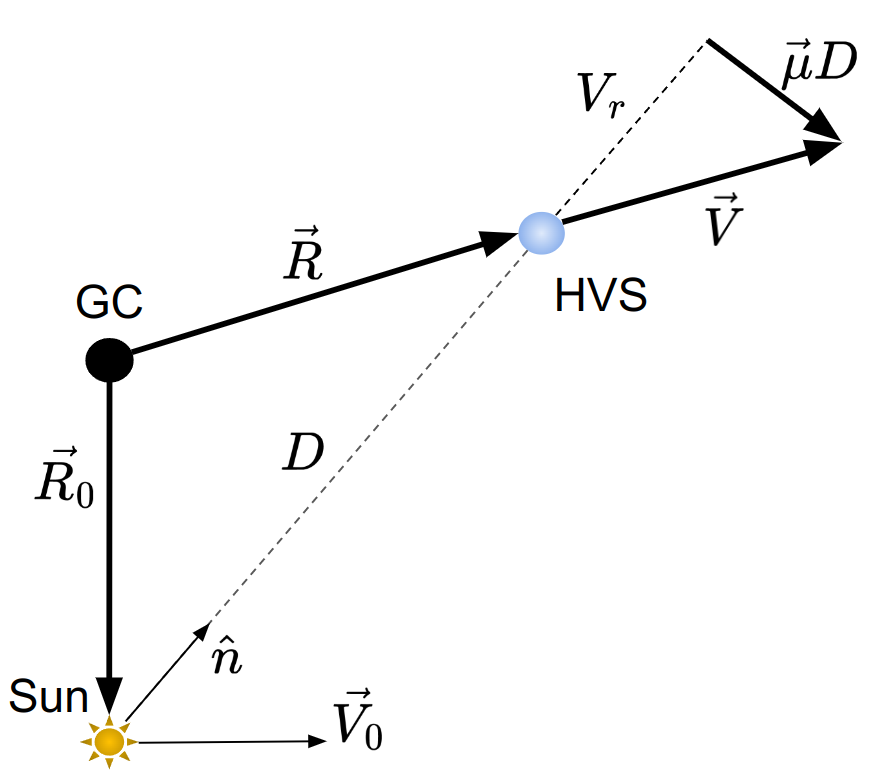}
    \caption{A diagram of the position ($\vec{R}$) and velocity ($\vec{V}$) of an HVS relative to the GC. This can be expressed in terms of a combination of the star's observed position ($\hat{n}$) and proper motion ($\vec{\mu}$) and measurements of the Sun's position relative to the GC ($\vec{R_0}$) and velocity ($\vec{V_0}$) along with the distance to the star ($D$) and its radial velocity ($V_r$).}
    \label{fig:overview}
\end{figure}

Let $\vec{R}$ be the vector from the GC to the star and $\vec{V}$ the star's velocity vector relative to the GC. We can write these down in terms of the observables 
\begin{itemize}
    \item $\vec{R_0}$: Vector from the GC to the Sun
    \item $\hat{n}$: Unit vector from the Sun to the source
    \item $D$: Distance from the Sun to the source
    \item $\vec{V_0}$: Velocity of the Sun in the Galactic frame of rest
    \item $V_{\mathrm{r}}$: Radial velocity of the source relative to the Sun
    \item $\vec{\mu}$: Proper motion of the source in the Heliocentric frame (always $\perp$ to $\hat{n}$)
\end{itemize}
giving
\begin{equation}
\label{eq:Robs}
\vec{R}=\vec{R_0}+D\hat{n}
\end{equation}
and
\begin{equation}
\label{eq:Vobs}
\vec{V}=\vec{V_0}+V_{\rm r}\hat{n}+D\vec{\mu}.
\end{equation}
We provide a sketch of the configuration in Fig.~\ref{fig:overview}. 
We use the measurement of $\vec{R_0}$ from \citet{Gravity_2018}, which is $8.122$ kpc, and $\vec{V_0}$ from \citet{Drimmel_2018}, which is $[12.9, 245.6, 7.78]$ \kms. The unit vector $\hat{n}$ is set by the sky coordinates of a star which are provided by \gaia, as is the proper motion vector $\vec{\mu}$. For most sources in \gaia DR3, approximately 1.5 billion, we have measurements for all but $D$ and $V_{\rm r}$. It is worth noting here that \gaia does measure parallaxes, but only for nearby sources do these result in precise distance measurements. The way of circumventing this lack of knowledge of $D$ and $V_{\rm r}$ is to only focus on objects that have radial trajectories pointing out from the GC. As we will see later, this enables us to calculate the distance and radial velocity from proper motion and sky position alone. For a star on a radial trajectory from the GC, its velocity aligns with the vector pointing from the GC to the star. This means the position $\vec{R}$ and velocity $\vec{V}$ are parallel and thus obey
\begin{equation}
\vec{R} \times \vec{V}=\vec{0}.
\end{equation}
We can expand this using equations \ref{eq:Robs} and \ref{eq:Vobs} leaving the terms
\begin{equation}
\label{eq:cross}
\vec{R_0} \times \vec{V_0} + 
V_{{\rm r}} (\vec{R_0} \times \hat{n} ) +
D  (\vec{R_0} \times \vec{\mu} +
  \hat{n} \times \vec{V_0}) +
D^2 (\hat{n} \times \vec{\mu} )
= \vec{0}
\end{equation}
and we can pull out the distance and radial velocity by taking the dot products with $\hat{n}$ and $\vec{\mu}$ and rearranging.

Taking the dot product of equation \ref{eq:cross} with $\hat{n}$ reveals
\begin{equation}
D = \frac{\hat{n}\cdot\left( \vec{R_0}\times \vec{V_0}\right)}{\vec{R_0}\cdot\left( \hat{n}\times \vec{\mu}\right)} = \frac{V_0}{\mu}\left(\frac{\hat{n}\cdot\left( \hat{R}_0\times \hat{V}_0\right)}{\hat{R_0}\cdot\left( \hat{n}\times \hat{\mu}\right)}\right)
\end{equation}
and similarly the dot product of equation \ref{eq:cross} with $\vec{\mu}$ gives
\begin{equation}
\begin{aligned}
V_{{\rm r}} &=  \frac{D\vec{V_0}\cdot\left( \hat{n}\times \vec{\mu}\right) - \vec{\mu}\cdot\left( \vec{R_0}\times \vec{V_0}\right)}{\vec{R_0}\cdot\left( \hat{n}\times \vec{\mu}\right)} \\
&= V_0  \left(\frac{\frac{D}{R_0} \hat{V}_0\cdot\left( \hat{n}\times \hat{\mu}\right) - \hat{\mu}\cdot\left( \hat{R}_0\times \hat{V}_0\right)}{\hat{R}_0\cdot\left( \hat{n}\times \hat{\mu}\right)}\right).
\end{aligned}
\end{equation}
This means that for a star moving on a radial trajectory, the distance and radial velocity are set by the observables available for $\sim1.5$B stars in \gaia DR3. 

For HVSs, the assumption that their trajectory is purely radial can be reasonable, since stars moving much faster than the escape velocity are not deflected significantly by the Galactic potential \citep{Kenyon_2018, Boubert_2020}. In this work, we apply the above presented method to determine the distance and radial velocity to all stars in \gaia DR3. Since those solutions are only physical for stars on radial trajectories, which most stars in \gaia are not, we refer to them as the implied distance, $D_{\rm I}$, and implied radial velocity, $V_{\rm r, I}$.

We convert $D_{\rm I}$ to implied parallax ($\varpi_{\rm I}$) to allow for convenient comparison to measurements by taking the inverse of $D_{\rm I}$. We calculate the uncertainty on $\varpi_{\rm I}$ assuming that the proper motion measurement is the only source of uncertainty. We make the simplifying assumption that the proper motion uncertainties are uncorrelated and since $\varpi_{\rm I}\propto\mu$, we can approximate the uncertainty on $\varpi_{\rm I}$ as
\begin{equation}
\sigma_{\varpi_{\rm I}} = \frac{\sqrt{\left[\vec{R}_0\cdot\left(\hat{n}\times\hat{\mu}_\delta \right)\sigma_{\mu_\delta} \right]^2 + \left[\vec{R_0} \cdot\left(\hat{n}\times\hat{\mu}_{\alpha*}\right)\sigma_{\mu_{\alpha}}\right]^2}}{\hat{n}\cdot\left(\vec{R}_0\times\hat{V}_0\right)},
\end{equation}
with $\hat{\mu}{_{\delta}}$ and $\hat{\mu}_{\alpha}$ the basis vectors for the proper motion in declination (Dec) and right ascension (RA) respectively, and $\sigma_{\mu_\delta}$ and $\sigma_{\mu_{\alpha}}$ the uncertainties on the proper motions in Dec and RA respectively. For a more accurate uncertainty analysis one would sample over the uncertainty on all observables.

\subsection{HVS candidate selection criteria}
\label{sec:selection_function}

The $\varpi_{\rm I}$ and $V_{\rm r, I}$ we derived in the previous section can be used to identify HVS candidates. Here we give a short conceptual overview of our approach. We first compare the implied parallax to the \gaia parallax and reject candidates for which they are inconsistent. In addition, we determine the location in the Hertzprung-Russel (HR) diagram according to the implied distance and remove any candidates for which we consider this solution to be unphysical. Furthermore, we only look at HVS candidates with a high enough 3D implied velocity for our radial trajectory assumption to be reasonable. Lastly, we apply additional cuts that are aimed to reduce the contamination of our sample. We base some of these cuts on catalogues of mock HVSs obtained from simulations described in Section~\ref{sec:simulations}. These simulations model the orbits of ejected HVS populations and use the characteristics of \gaia to predict their observational properties.

Having discussed the general principle of how we select HVS candidates in \gaia, we now provide the detailed requirements set for a particular star to be observed in our survey. We categorise these into four groups, each discussed in detail below and summarised in Table~\ref{tab:selections}. The goal of the selections presented here is to increase the purity of the sample; i.e.~reduce the number of non-HVSs while retaining the real HVSs as much as possible. This is motivated by the necessarily limited number of sources we can provide follow-up observations for.

\subsubsection{Astrometric and kinematic selections}
\label{sec:astro_kine}
We start by using the astrometric and kinematic properties of stars to select potential HVSs. HVSs need to be very fast to prevent significant deviation from their otherwise radial trajectories by the torque on the orbit generated by the non-sphericity of the Galactic potential. For this reason, we only consider stars with total implied velocity, $V_{\text{I}}$, in the range $[800, 3500]$ \kms, independent on Galactocentric distance. The lower limit ensures that (excluding the inner GC) all stars would be unbound. The upper limit includes most of the extremely fast predicted HVSs \citep{Rossi_2014}, while limiting contamination by non-HVSs (which can have extremely high implied velocities). Secondly, we compare the implied parallax to the parallax measured by \gaia. We only select sources with implied parallax consistent with measured parallax within $2\sigma$, considering both the uncertainty on the measured and the implied parallaxes. We do this comparison to the parallax and not the distance since the parallax has Gaussian uncertainties.

The source of uncertainty on $\varpi_{\text{I}}$ is the measured uncertainty on $\vec{\mu}$ from \gaia. Because comparing parallaxes is not constraining when both are highly uncertain, we only consider sources where $\varpi_{\text{I}}$ over its associated uncertainty is larger than five. In addition, most parallaxes in the \gaia catalogue have large fractional uncertainties. This also means that requiring the implied parallax and measured parallax to be consistent is often not a stringent requirement. To alleviate this, we additionally compare the implied distances to the photo-geometric ones determined in \citet{Bailer-Jones_2021}. The photo-geometric distances make use of the colour-magnitude information from stars by incorporating a colour-magnitude dependent prior on the extinction corrected absolute magnitude, which varies as a function of sky position. Similar to the parallax selection, we require that the implied distances are consistent with the photo-geometric ones. As a threshold we use $2\sigma$, where we use the 16th and 84th percentiles from \citet{Bailer-Jones_2021} as the negative and positive $1\sigma$ estimates for the posterior respectively.

\subsubsection{Photometric selections}
\label{sec:photo_selections}
The implied position of many stars in the HR diagram (according to the implied distance) will be unphysical, because the implied distances are only appropriate for stars moving on radial trajectories. To remove unphysical solutions, we only consider sources within part of the HR diagram. 

We first correct the G, ${\rm G_{RP}}$, and ${\rm G_{BP}}$ magnitudes for extinction using the 2D extinction map from \citet{Schlegel_1998} and the recalibration from \citet{Schlafly_2011}, assuming every source is behind the extinction layer. We use an $R_{\rm V}=A_{\rm V}/{\rm E(B-V)}$ of 3.1 with the \citet{Fitzpatrick_1999} extinction law. We use the values provided by \gaia to convert the extinction at 550 nm to the relevant bands\footnote{\url{https://www.cosmos.esa.int/web/gaia/edr3-extinction-law}}. Since those corrections are formally only valid when the intrinsic colour is known, we perform 10 iterations over the extinction coefficients, adjusting the colour of the source at each step. We only select sources which are implied to be within $\Delta {\rm M_G}=2$ of the main-sequence, which we define to be ${\rm M_G} = 4.3\times({\rm G_{BP}-G_{RP}}) + 0.5$. This is a fairly simplistic selection, which can be improved upon in future surveys by, for instance, taking the relative contamination of our HVS candidate catalogue by field stars into account as a function of position on the HR diagram. To exclude highly extincted regions, we only consider HVS candidates with a G-band extinction coefficient $A_{\text{G}}<1.5$.

\begin{figure}
    \centering
    \includegraphics[width=\linewidth]{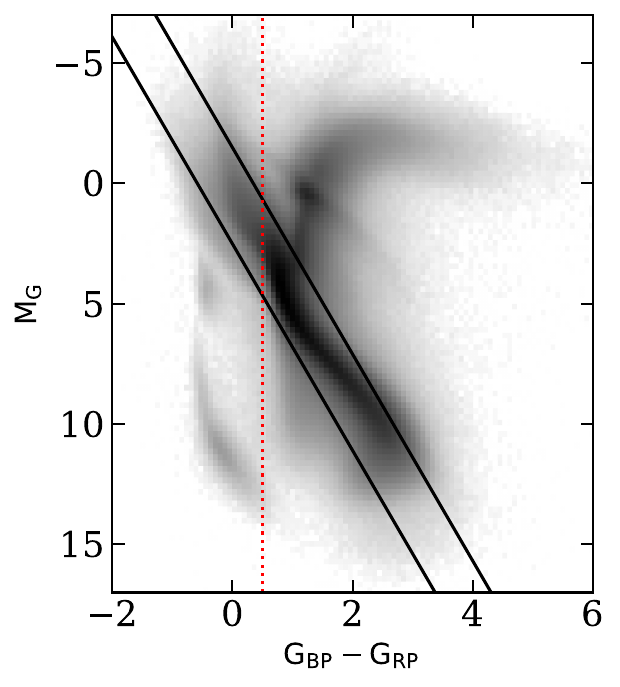}
    \caption{Reference HR diagram created from sources in \gaia. The dotted line gives our colour limit, only stars to the left of which we consider. The solid lines indicate the main-sequence region within which we consider HVS candidates. The colour scale is logarithmic.}
    \label{fig:HRD}
\end{figure}

We noticed that the relative contamination of our HVS candidate catalogue by field stars was lowest for blue sources, based on the simulations we discuss in Section~\ref{sec:model}. For this reason we select only sources with an extinction corrected \bprp colour below 0.5, which corresponds to a mass $\gtrsim1.3$ M$_\odot$. We show these selections in Fig.~\ref{fig:HRD} in combination with a reference HR diagram for sources in \gaia.
To provide perspective, we also overlay the density of HVS candidates in \gaia in Fig.~\ref{fig:app_HR} in the appendix.

\subsubsection{Sky coordinate selections}
\label{sec:coordinate_selections}
We know that the distribution of HVSs on the sky should be anisotropic, even for an isotropic ejection mechanism in the GC due to on-sky projection. Moreover, we expect that the contamination of our HVS candidate sample is highest near the GC and anti-centre, where our selections are relatively ineffective. In addition, in the Galactic plane we expect the contamination to be higher than at high Galactic latitudes due to the high relative number of field stars to expected HVSs. In order to limit contamination by field stars we define the expected number of HVSs to number of candidates for different directions in the sky. For this purpose we use \textit{HEALPix}\footnote{\url{http://healpix.sf.net/}} \citep[$nside = 3$;][]{Gorski_2005, Zonca_2019} to divide the sky into equal area pixels. We determine the expected density of HVSs across the sky using the simulations described in Section~\ref{sec:model}. We only observe sources in \textit{HEALPix} pixels where the ratio of the density of expected HVSs to candidates (normalised by the number of candidates) is larger than five\footnote{The \texttt{HEALPix} pixels we use are 0-19, 22-31, 35-42, 46-55, 58-62, 65, 66, 69-73, 78, 82-85, 93-96, 101, 102, 103 and 107 in the ring scheme based on RA and Dec.}.

To prevent contamination from stars in the Large Magellanic Cloud (LMC) or Small Magellanic Cloud (SMC), we additionally require any HVS candidate to be more than 8 and 3 deg from their respective centres\footnote{The LMC centre we use has $l=280.4652$, $b=-32.8884$ deg. The SMC centre we define to be at $l=302.8084$, $b=-44.3277$ deg.}. 

The selection steps up to this point leave us with only 600 HVS candidates, the previously identified S5-HVS1 being one of the top candidates \citep{Koposov_2020}. We provide an overview of the selections in Table~\ref{tab:selections}. 
\begin{table}
    \centering
    \caption{Overview of the HVS candidate selections used in this study.}
    \begin{tabular}{lc}
    \hline
         Selection & Section for reference\\
         \hline
         $\varpi_{\rm I} - \varpi_{\rm Gaia} < 2\sigma$ & \ref{sec:astro_kine}\\
         $800 < V_{\rm I} < 3500$ \kms & \ref{sec:astro_kine}\\
         $\varpi_{\rm I}/\sigma_{\varpi_{\rm I}} > 5$ &\ref{sec:astro_kine}\\
         RUWE $<1.4$ & \ref{sec:astro_kine}\\
         $D_{\rm I} - D_{\rm BJ} < 2\sigma$ & \ref{sec:astro_kine}\\
         $-1.5 <M_{\rm G} - 4.3\times({\rm G_{BP}-G_{RP}}) < 2.5$ & \ref{sec:photo_selections}\\
         $A_{\rm G} < 1.5$ & \ref{sec:photo_selections}\\
         ${\rm G_{BP}-G_{RP}} < 0.5$ & \ref{sec:photo_selections}\\ 
         $N_{\rm HVS, sim}/N_{\rm HVS, candidates} > 5$ & \ref{sec:coordinate_selections}\\
         Separation from LMC $>8$deg & \ref{sec:coordinate_selections}\\
         Separation from SMC $>3$deg & \ref{sec:coordinate_selections}\\
         \hline
    \end{tabular}
    \label{tab:selections}
\end{table}
The complete source list can be found online\footnote{\url{https://zenodo.org/doi/10.5281/zenodo.12179452}}. We provide the first five sources in Table~\ref{tab:source_table}.
\begin{table}
    \centering
    \caption{Here we provide the first five sources, ordered by {\tt source\_id}, from our catalogue of 600 HVS candidates with their \gaia DR3 {\tt source\_id}, implied heliocentric radial velocity, and implied heliocentric distance.}
    \begin{tabular}{l|c|c}
         \hline
         \gaia DR3 {\tt source\_id} & $V_{\rm r, I}$ [\kms]& $D_{\rm I} [kpc]$ \\
         \hline
         16647293239608960 & 805 & 6.6 \\
         57537894455885184 & 1429 & 15.2 \\
         292010016092347648 & 829 & 21.1 \\
         307220183209400448 & 638 & 12.2 \\
         331500110074917120 & 1092 & 18.7 \\
         ... &&\\
         \hline
    \end{tabular}
    \label{tab:source_table}
\end{table}

\subsubsection{Instrument-specific selections}
\label{sec:instrument_selections}
We used two observatories for follow-up observations of our candidates; the Isaac Newton Telescope (\INT) and the New Technology Telescope (\NTT). The \INT is a 2.54-meter telescope located on the island of La Palma and the \NTT is a 3.58-meter telescope sited at La Silla in Chile. We used two observatories to gain both northern and southern hemisphere coverage of the sky.
The instrument-specific selection for the \INT is 
\begin{itemize}
    \item $\textrm{G}<19$
    \item $\textrm{Dec} > -20$ deg
    \item $\textrm{RA} < 60$ deg or $\textrm{RA} > 230$ deg.
\end{itemize}
Our instrument specific selection for the \NTT is 
\begin{itemize}
    \item $\textrm{G}<19.3$
    \item $\textrm{Dec} < 0$ deg.
\end{itemize}
We find a total of 284 HVS candidates in \gaia that match all of the listed selections. We performed spectroscopic follow-up observations for these sources, which are described in Section~\ref{sec:observations}. 

\section{Observations}
\label{sec:observations}
In the previous section, we discussed how we select our HVS candidates. In this section we describe the follow-up observations we performed for those candidates. We describe the instrument set up in Section~\ref{sec:set_up}, the data reduction in Section~\ref{sec:data_reduction}, the spectral analysis in Section~\ref{sec:spectral_analysis}, and lastly the observed stars in Section~\ref{sec:observed}. 

An overview of our observations is given in Table~\ref{tab:obs}.
\begin{table}
    \centering
    \caption{Overview of the observing dates for our survey.}
    \begin{tabular}{l|l|l}
        \hline
        Instrument & Observing dates & Run number \\
        \hline
        INT & 2022 Aug 20-21 & ING.NL.22B.002\\
        INT & 2022 Sep 19-30 & ING.NL.22B.002 \\
        NTT & 2022 Nov 25-28 & 110.23SU.001 \& 110.23SU.002\\
        NTT & 2023 June 3 & 111.24MP.001 \\
        NTT & 2023 July 14-15 & 111.24MP.002\\
        \hline
    \end{tabular}
    \label{tab:obs}
\end{table}

\subsection{Instrument set up}
\label{sec:set_up}
To confirm or reject HVS candidates, we do not require high radial velocity precision. We therefore set up the instruments to allow efficient radial velocity measurements with uncertainties of a few tens of \kms. 

At the \INT, we use the Intermediate Dispersion Spectrograph (IDS) with the EEV10 detector and the R400V grating. This gives us an approximate resolution of $\textrm{R} = \frac{\lambda}{\Delta\lambda} = 1600$ at 4500 \AA \footnote{\url{https://www.ing.iac.es/astronomy/instruments/ids/idsgrat_tables.html}}. We were allocated a total of twelve nights of observing at the \INT.

At the \NTT, we use the ESO Faint Object Spectrograph and Camera (v.2; EFOSC2) with grism Gr\#07. This setup gives us a resolution of about 500 at 3800 \AA\footnote{\url{https://www.eso.org/sci/facilities/lasilla/instruments/efosc/inst/Efosc2Grisms.html}}. We were allocated seven nights of observing at the \NTT.

\subsection{Data reduction}
\label{sec:data_reduction}
In the following, we describe the data reduction steps applied to the raw data from the \INT and \NTT. We use the same pipeline for the two telescopes, with minor adjustments where needed to account for the different instruments.\

We perform standard bias subtraction and flat-field correction to all science spectra. We then mask out any cosmic rays using \texttt{ccdproc} \citep{matt_craig_2022_6533213}. We trace and subsequently extract the spectra using \texttt{PyRAF}, which is the Python implementation of \texttt{IRAF} \citep{Doug_1986, Dough_1993}. Background subtraction is performed by selecting a region around the science spectrum. Through visual inspection, we make sure there are no sources in the background region and make adjustments where needed. In the case of poor seeing during the observation, for instance, the background region might need to be further separated from the centre of the science spectrum. In the case of the \INT, wavelength calibration is performed using the CuAr+CuNe lamp. For the \NTT we make use of the He and Ar lamps. An arc spectrum is taken after every science exposure to account for instrument flexures. No flux calibration is applied, since it is not required for radial velocity measurements.

We additionally calculate the signal-to-noise (S/N) as a function of wavelength. We start by taking the square root of the sum of squares of the photon noise and the readout noise. We remark that this is an approximation, because in the regime of low photon counts, the constraint on the flux is not Gaussian \citep[see e.g.][section 4.2.9]{Guy_2023}. We use the same trace used in extracting the science spectrum to extract the noise spectrum and perform error propagation in subtracting the background.

\subsection{Spectral analysis}
\label{sec:spectral_analysis}
Having obtained wavelength calibrated spectra, we now describe our analysis. We determine the radial velocities from our reduced data using \texttt{RVSpecFit} \citep{Koposov_2019}. This software relies on direct pixel fitting using interpolated stellar templates from the PHOENIX spectral library \citep{Husser_2013}. For the spectra from the \NTT we use the wavelength interval $3500 < \lambda < 5240$ \AA, while for the \INT we use the wavelength interval $3600 < \lambda < 6800$ \AA. We perform barycentric correction on the resulting radial velocities using \texttt{Astropy}.

\subsection{Observed HVS candidates}
\label{sec:observed}
\begin{figure}
    \centering
    \includegraphics[width=\linewidth]{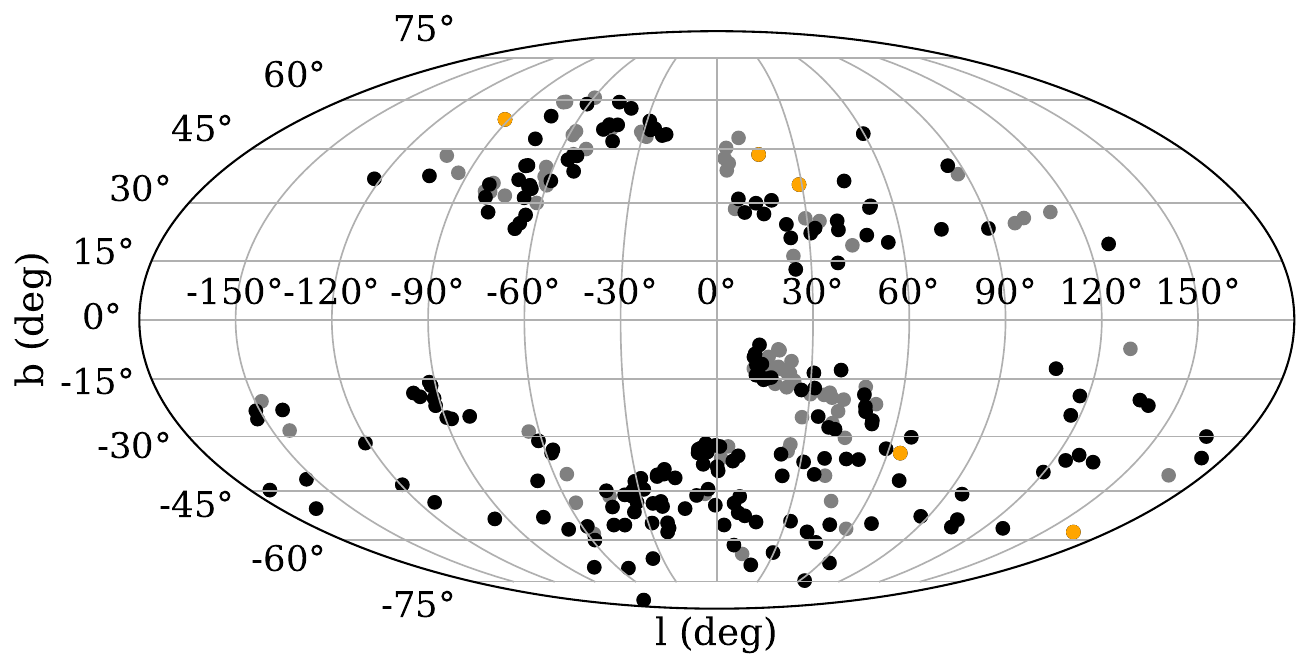}
    \caption{Sky distribution of our HVS candidate sample in Galactic coordinates. Black points indicate the stars with follow-up observations and the orange points the ones with either {\it SDSS} or {\it LAMOST} radial velocity measurements. For the grey stars we do not have radial velocity measurements.}
    %\emr{a reliable one? i.e., low signal to noise spectrum or no follow up at all?}.}
    \label{fig:sky_distr}
\end{figure}

We obtained radial velocity measurements for 196 out of our 284 HVS candidates. This translates to a completeness factor of about 69\%. We only consider spectra with a S/N larger than two and three for the \INT and \NTT respectively. For lower S/N spectra we were not able to recover reliable radial velocity measurements.
%\emr{say precisely for how many stars}. 
For another five stars, we found previous radial velocity measurements in the {\it SDSS} DR14 low-resolution stellar catalogue \citep{Abolfathi_2018} or in the {\it LAMOST} DR8 low-resolution catalogue \citep{LAMOST_2012}. We give the data on these stars in the appendix Table~\ref{tab:external}. We will also consider these five sources in our observational sample, bringing our total completeness to about 71\%. In Fig.~\ref{fig:sky_distr} we plot the sky distribution of our HVS candidate catalogue, indicating which ones have been observed successfully.
We provide an observational log of our observations in an online table\footnote{\url{https://zenodo.org/doi/10.5281/zenodo.12179452}}, the first five rows of which are shown in Table~\ref{tab:obs_results}.
\begin{table*}
    \centering
    \caption{Five sources from our observed HVS candidates in no particular order. The full table can be accessed online (see main text). The radial velocities are in the barycentric frame.}
    \begin{tabular}{l|c|c|c|c}
         \hline
         \gaia DR3 {\tt source\_id} & radial velocity [\kms] & radial velocity uncertainty [\kms] & date of observation & instrument \\
         \hline 
         6408940170344856320 & 220.6 & 38.0 & 2022-11-28 & NTT\\
         16647293239608960 & 42.4 & 40.6 & 2022-09-23 & INT\\
         4080886923180284928 & 32.8 & 35.4 & 2023-06-04 & NTT\\
         3755045965083116160 & 281.0 & 21.2 & 2022-11-26 & NTT\\
         4195180362312873984 & -75.7 & 80.2 & 2023-07-16 & NTT\\
         ...&&&&\\
         \hline          
    \end{tabular}
    \label{tab:obs_results}
\end{table*}

\begin{figure}
    \centering
    \includegraphics[width=\linewidth]{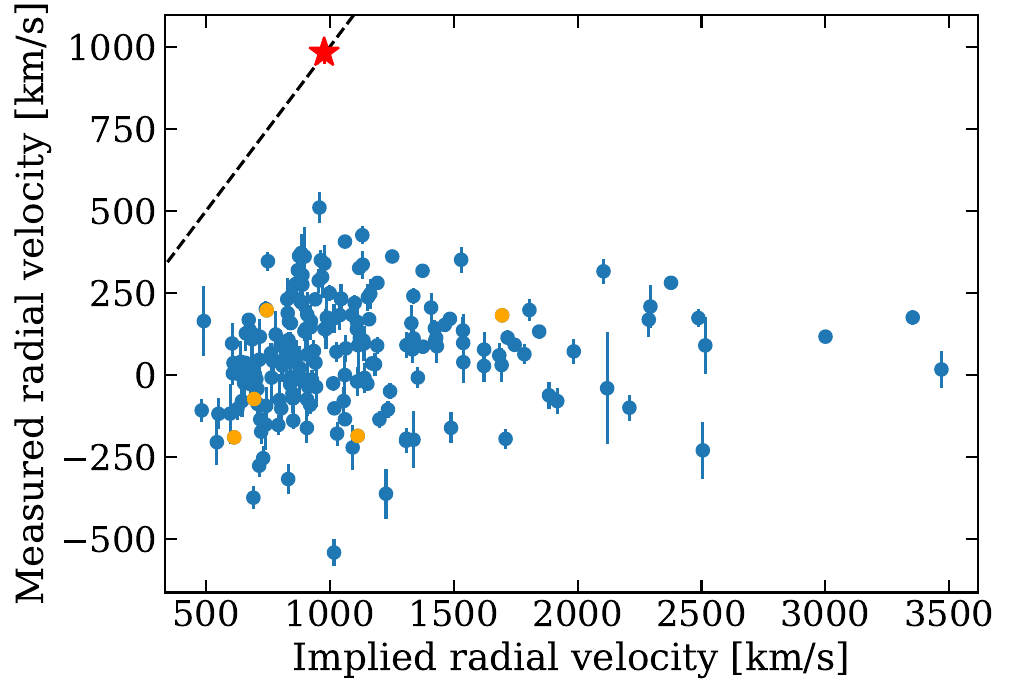}
    \caption{Implied radial velocity against the measured radial velocity for the HVS candidates we observed. The dashed line gives the bisection. The blue points indicate our measurements, the red star highlights our measurement of S5-HVS1, and the orange points show the stars with radial velocity measurements from {\it SDSS} and {\it LAMOST}.}
    \label{fig:Vi_vs_Vr}
\end{figure}

In Fig.~\ref{fig:Vi_vs_Vr} we plot the implied radial velocity against the measured radial velocity.
We can see that only the previously identified S5-HVS1 falls on the bisection, clearly identifying it as an HVS. However, not all HVSs will be on the bisection, because either their trajectories are not purely radial or due to observational uncertainty. In practice, this means we can not state with absolute certainty that the other observed sources are not HVSs, because of observational uncertainty and orbits which are not purely radial. We will return to this point in Section~\ref{sec:selection_simulations}.

\section{HVS mock catalogues}
\label{sec:model}
In this work we use a comparison between observations and simulations to infer properties of HVSs. Section~\ref{sec:Hills} describes the HVS ejection mechanism and Section~\ref{sec:simulations} discusses the simulation suite in which it is applied. We use these simulations to make informed decisions for our HVS candidate selection, mentioned in Section~\ref{sec:selection_function}. Most importantly, the simulations allow us to provide constraints on the ejection of HVSs and the GC environment as presented in Section~\ref{sec:constraints}.

\subsection{Hills mechanism}
\label{sec:Hills}
We assume HVSs are ejected through the Hills mechanism proposed in \citet{Hills_1988} \citep[for a review see][]{Brown_2015}. In this mechanism, a binary star system approaches a massive black hole (MBH) within the tidal radius, where the tidal force from the MBH exceeds the binary self gravity. The result is that the binary is separated; one star is captured by the MBH in a tight, elliptical orbit and the other star is ejected as an HVS. If the progenitor binary is on a parabolic orbit, which is an excellent approximation in the scenario assumed here \citep{Kobayashi_2012}, both stars in the binary have an equal likelihood of being ejected, independent of mass ratio \citep{Sari_2010}. The velocity at which the HVS is ejected from the sphere of influence of the MBH is given by
\begin{equation}
\label{eq:vej}
V_{\text{ej}} = \sqrt{\frac{2Gm_{\text{c}}}{a}}\left(\frac{M_{\text{MBH}}}{M}\right)^{1/6},
\end{equation}
where we neglected a multiplicative constant factor of order unity which depends on the geometry of the encounter \citep[see, e.g., Fig.10 in][]{Sari_2010}. In equation \ref{eq:vej}, $m_{\text{c}}$ is the mass of the captured companion, $M_{\text{MBH}}$ in the case of the Galaxy is the mass of Sgr A$^*$ \citep[$4\times10^6$ M$_\odot$,][]{Eisenhauer_2005, Ghez_2008, Schodel_2009, Gillessen_2009, Boehle_2016, Gillessen_2017, Do_2019, Gravity_2019}, and $M$ the total mass of the progenitor binary \citep{Sari_2010, Kobayashi_2012, Rossi_2014}. Most of the stars ejected through this mechanism are still bound to the Galaxy, but the ejections can occur at velocities in excess of 1000 \kms, putting stars on orbits unbound to the Galaxy.

\subsection{Simulations}
\label{sec:simulations}
The simulation framework we use is \texttt{Speedystar}\footnote{\url{https://github.com/fraserevans/speedystar}} \citep{Contigiani_2019, Evans_2022_II}. This code is able to create mock observational catalogues of Hills mechanism HVSs, and of stars ejected by a binary massive black hole \citep{Evans_2023}. Here we focus exclusively on the Hills mechanism. Below we describe the most important steps to understand the model, for details we refer to \citet{Evans_2022_II}. 

We start by creating an HVS progenitor population of binaries. These binaries are characterised by the primary zero-age main sequence mass ($m_{\text{p}}$), mass ratio ($q$), and the orbital semi-major axis ($a$) at the moment of disruption. The mass of the primary is drawn from a mass function (MF) which follows a power law distribution of the form $f(m_{\text{p}})\propto m_{\text{p}}^{-\kappa}$ in the range $[0.1, 100]$ M$_\odot$. The secondary is then assigned a mass according to a mass ratio distribution of the form $f(q)\propto q^\gamma$, with $0.1 \leq q \leq 1$. Lastly, the orbital semi-major axis are assigned assuming a binary orbital period distribution of the form $f(\log P) \propto (\log P)^\pi$. Different values for the three parameters $\kappa$, $\gamma$, and $\pi$ are explored with uniform priors to account for uncertainty in the HVS progenitor population. For $\kappa$ we use the interval $[0.3, 3.3]$ to allow for both very top-heavy and bottom-heavy MFs. For $\gamma$ we use the interval $[-2, 2]$ and for $\pi$ the interval $[0, 2]$. We allow for the wide range in MF index because of the uncertainty on the IMF near the GC, alluded to in the introduction. IMF indexes of 0.45, 1.7, and a canonical 2.3 have all been claimed near Sgr A$^*$ \citep[][respectively]{Bartko_2010, Lu_2013, Lockmann_2010}. The parameter range on $\pi$ allows for a log-uniform period distribution (also known as \"Opik's law), or distributions with increasing fractions of wide binaries. The parameter range on $q$ encapsulates the results from different studies into massive binaries in star forming regions \citep{Sana_2012, Sana_2013, Moe_2013, Moe_2015, Dunstall_2015}.

For each star we assume solar metallicity, with $Z_\odot=0.0142$ \citep{Asplund_2009}. Additionally, we performed tests with super-solar metallicity values of $[\mathrm{Fe/H}]=0.3$, since stars near the GC are mostly metal rich \citep{Do_2015, Feldmeier_2017, Nandakumar_2018, Schultheis_2019}.

From this progenitor population, binaries are drawn at a constant rate. One of the binary components is selected at random and assigned an ejection speed according to equation~\ref{eq:vej}. We assume the ejection is isotropic and assign the star a random age within the total lifetime of the shorter-lived star of the progenitor binary, where we implicitly assume a constant star formation rate. The ejected stars are subsequently integrated in the potential of the Galaxy \citep[see][table 1]{Marchetti_2018}. The result is a simulated current population of HVSs throughout the Galaxy.  These HVSs are evolved using single stellar evolution models from \citet{Hurley_2000} in {\tt AMUSE} \citep{PORTEGIESZWART_2009, PORTEGIESZWART_2013, Pelupessy_2013, PORTEGIESZWART_2018}.

Lastly, mock observations are performed incorporating the visual extinction along the line of sight, effective temperature, surface gravity, and metallicity of the simulated HVSs. This is done using the MESA Isochrone and Stellar Tracks model grid \citep{Dotter_2016, Choi_2016}. This allows us to determine the magnitude and colour of these stars as they would be observed by various observatories. Astrometric uncertainties in parallax and proper motion, as observed by \gaia, are determined using the DR3 astrometric spread function of \citet{Everall_2021}.

\subsection{HVS candidate selection applied to simulations}
\label{sec:selection_simulations}
Now that we have described both our simulations and observations, we will describe how we apply the selections discussed in Section~\ref{sec:selection} to our simulated HVSs.

Firstly we have to consider which stars are observed by \gaia in the first place. Although the magnitude limit of \gaia is $\text{G}=20.7$ \citep{Gaia_2016}, not all sources down to that depth appear in the published \gaia catalogue for various reasons. A description of which stars do and do not appear in the published \gaia catalogue is referred to as the \gaia selection function, which has been empirically determined in \citet{Cantat-Gaudin_2023}\footnote{\url{https://github.com/gaia-unlimited/gaiaunlimited?tab=readme-ov-file}}. Although \gaia is virtually complete down to our magnitude limit of ${\rm G}=19.3$ except in crowded regions, we nonetheless incorporate the empirical \gaia selection function for completeness. We also convolve the `true' parallaxes and proper motions with their uncertainties. No uncertainties are considered on the photometric measurements.

Most of the selections described in Section~\ref{sec:selection_function} can be directly applied to the resulting mock HVS catalogue. We can calculate the implied distance and radial velocity for the mock HVSs in the same way we did for the \gaia catalogue and apply our selections. The same can be done for the photometric, sky coordinate, and instrument-specific selections. An exception is the photo-geometric distance from \citet{Bailer-Jones_2021}. To apply this selection to the mock HVS catalogue, we resort to the \citet{Bailer-Jones_2015} source code (C.A.L.~Bailer-Jones, private communication) and calculate the photo-geometric distances for the simulated HVSs. After obtaining the photo-geometric distances for all our mock HVSs in this way, we again apply the same selections to our mock catalogue as to the \gaia catalogue. The only exception is the {\tt RUWE}, which we do not model for our simulated HVSs. We expect HVSs to be single stars and for single stars fewer than one in a million have ${\tt RUWE}>1.4$ \citep{Penoyre_2022}.

\begin{figure}
    \centering
    \includegraphics[width=\linewidth]{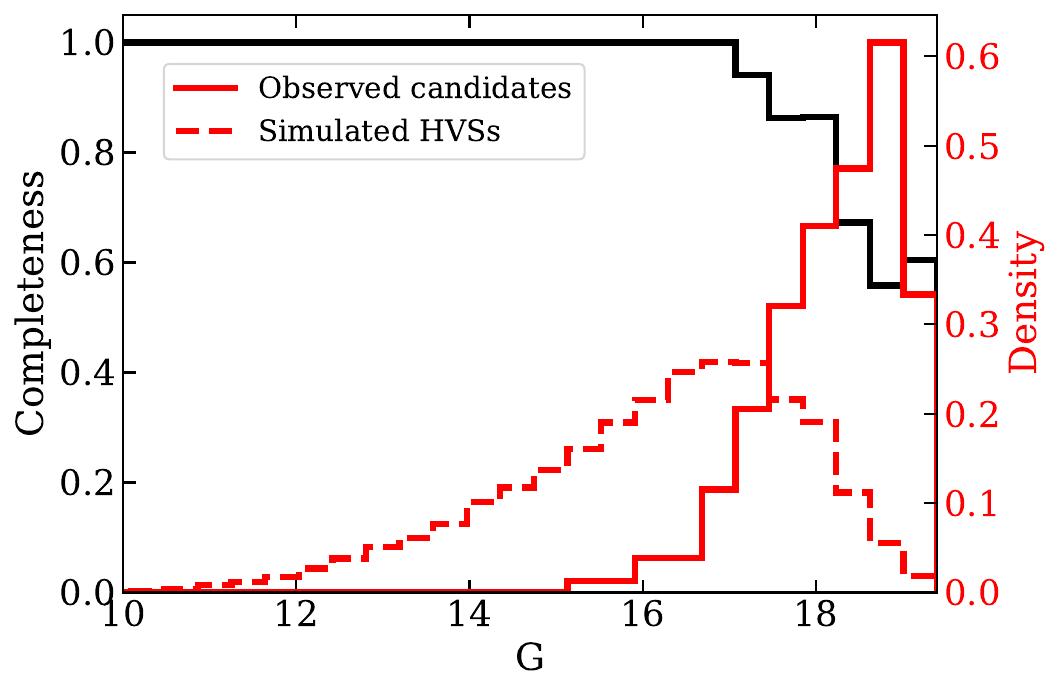}
    \caption{Observational completeness of our sample of HVS candidates as a function of apparent G magnitude. We also show the distribution of simulated HVSs and our observed candidates in G magnitude in red on the right axis.}
    \label{fig:compl_g}
\end{figure}

\begin{figure}
    \centering
    \includegraphics[width=\linewidth]{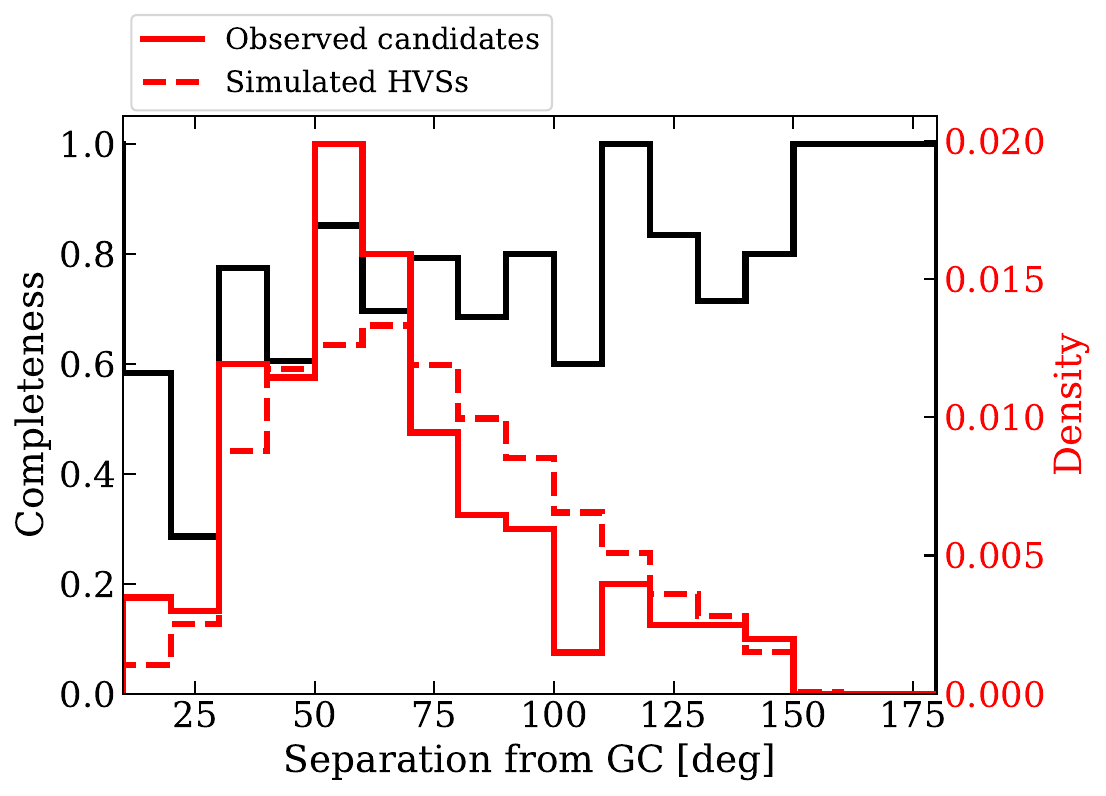}
    \caption{Observational completeness of our sample of HVS candidates as a function of the angle to the GC. In addition, we show the distribution of simulated HVSs and our observed candidates as a function of their angular separation from the GC in red on the right axis.}
    \label{fig:compl_gc_ang}
\end{figure}

Since our observational catalogue does not cover every candidate from the selections introduced in Section~\ref{sec:selection}, we need a prescription to quantify our observational completeness. In Fig.~\ref{fig:compl_g} we show our follow-up observational completeness as a function of the apparent magnitude of the HVS candidates.
For each bin in G magnitude, we calculate the fraction of sources we observed over the number of sources in our catalogue (see Section~\ref{sec:instrument_selections}). As can be seen, the observational completeness decreases for fainter sources, since we prioritised bright sources. In addition, we have lower completeness for regions with many HVS candidates that were not accessible for a significant amount of time during the observational runs. This expresses itself mainly in a lower completeness near the GC. In Fig.~\ref{fig:compl_gc_ang} we show our follow-up observational completeness as a function of the angle to the GC.
Since the observational completeness shows a stronger bias in G-magnitude than in the sky position, we decide to model the observational completeness only on G-magnitude. We bin the magnitudes of our candidates using 50 bins between ${\rm G}=0$ and ${\rm G}=19.4$ and calculate the fraction of observed stars within each bin. If there are no candidates in a bin, we consider the observations to be complete. The model of the selection function of our observations is then the completeness fraction per bin in G-magnitude, which we can apply to our simulations.

Lastly, we noted in Section~\ref{sec:observed} that we cannot reject with absolute certainty the HVS candidates  that have significantly different observed and implied radial velocities. To account for this, we only consider simulated HVSs for which $V_{{\rm r,I}}-V_{{\rm r}} < 300$ \kms, which holds for about 86\% of HVSs in our simulations. None of the sources we observed, except S5-HVS1, meet this criterion.

By combining all observational selections and applying them to the simulations, we can predict the number of HVSs that should be in our observational sample, which we use to provide constraints on the GC in Section~\ref{sec:HVS_parameter_constraints}.

\section{Constraints on the Galactic Centre}
\label{sec:constraints}
Having described the observations, we can now provide physical constraints on the GC environment and HVS ejection process. Before we do so in Section~\ref{sec:HVS_parameter_constraints}, we will first discuss the properties of HVSs that can be observed with \gaia in Section~\ref{sec:HVS_observable_by_gaia}.\

\subsection{HVSs observable by \gaia}
\label{sec:HVS_observable_by_gaia}
Not all HVSs can be observed by \gaia. Low-mass stars, for instance, can only be observed out to short heliocentric distances, due to their intrinsic low brightness. To establish the properties of HVSs that can be observed by \gaia, we make use of the simulations described in Section~\ref{sec:simulations}. We require that to be observed by \gaia and be considered an HVS, the star must have an apparent G magnitude below $20.7$, in addition to travelling on an unbound orbit. We additionally consider the \gaia selection function mentioned in Section~\ref{sec:selection_simulations}.

\begin{figure*}
    \centering
    \includegraphics[width=\textwidth]{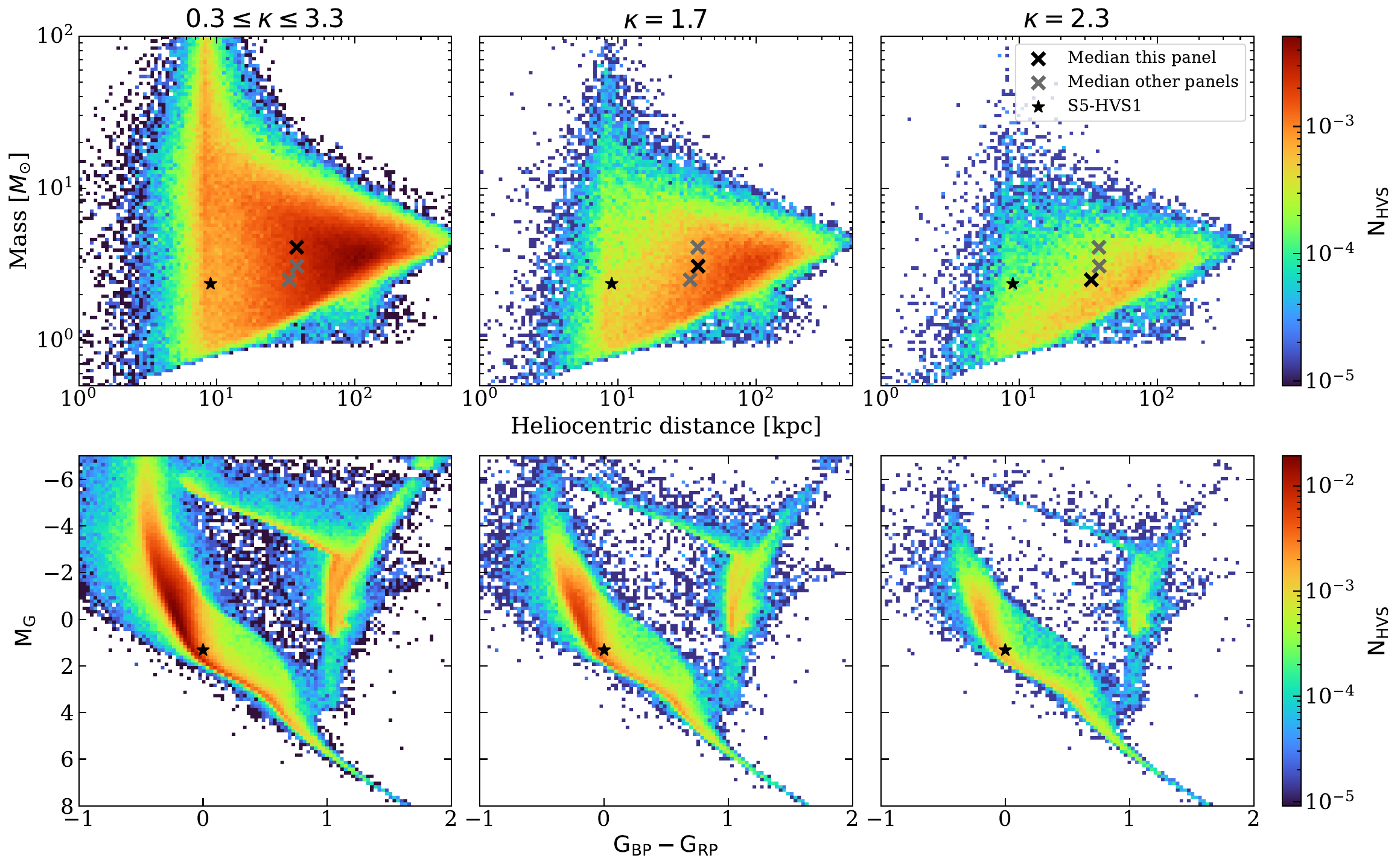}
    \caption{Predicted population of unbound HVSs in the \gaia catalogue. The three columns display different assumptions on the MF of the primary of the HVS progenitor binary population, as indicated above the respective columns. The numbers per bin are calculated for a fiducial ejection rate of $10^{-6}$ yr$^{-1}$. \textit{Top row}: the distance-mass distribution for predicted HVSs in \gaia. The black crosses indicate the median values of the respective panels, while the grey crosses indicate those for the other panels for easy comparison. The black star indicates the position of S5-HVS1. \textit{Bottom row}: the extinction corrected colour-magnitude distribution for predicted HVSs in \gaia.}
    \label{fig:observable_by_gaia}
\end{figure*}

For every set of model parameters we can now find the stars that would appear in the \gaia catalogue. We caution that due to observational uncertainties, not all of these real HVSs could be trivially identified as such; in practice, it can be challenging to determine if a star is in fact unbound or not. In Fig.~\ref{fig:observable_by_gaia} we plot the distance against the mass for HVSs that can be observed by \gaia, along with the colour-magnitude diagram.
In the left column, we add the contributions from all MFs within our prior range (see Section~\ref{sec:simulations}). In the middle and right columns we show the distribution for two specific MFs (a top-heavy one, $\kappa=1.7$, and one close to the canonical Salpeter MF, $\kappa=2.3$). The resulting distribution should be seen as a prediction for the overall population distribution of HVSs present in the \gaia catalogue, discovered or undiscovered. We can see that most of the HVSs in the \gaia catalogue will be at heliocentric distances between about 10 and 200 kpc, be on the order of a few solar masses, and be main-sequence stars. Below heliocentric distances of $\sim5$ kpc, the volume covered by \gaia is too small which makes discovering any HVSs within this region highly unlikely. The vertical feature in the top row at $\sim8$ kpc is the GC, where, due to projection effects, an over density of HVSs on the sky is present. The cutoff feature towards low-mass/distant stars is caused by the detection limit of \gaia in apparent magnitude. The stars below this boundary are giants, which are visible out to further distances than main-sequence stars. The upper right of the figure is not populated, because extremely massive stars do not live long enough to travel far from the GC. In Appendix Fig.~\ref{fig:app_G}, we additionally show the apparent magnitude distribution for am MF index $\kappa=2.3$.\

Interestingly, the MF has little effect on the ratio of evolved to main-sequence stars we expect to be within the \gaia catalogue. The fraction of evolved HVSs is expected to be about 0.17, changing by less than 0.03 between different MFs.

\begin{figure}
    \centering
    \includegraphics[width=\linewidth]{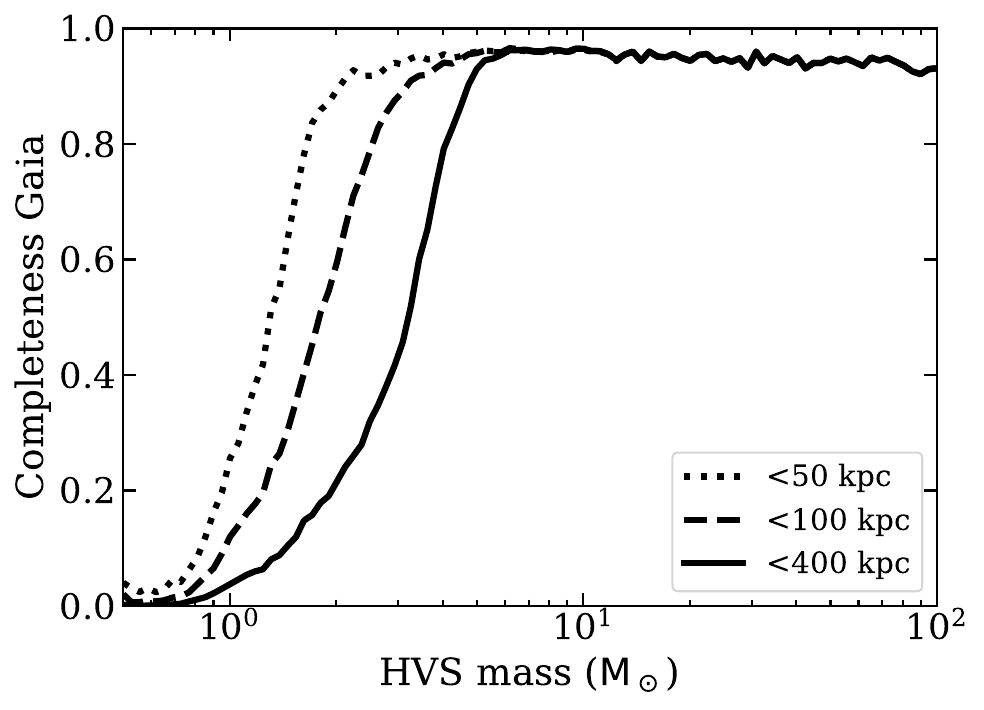}
    \caption{Completeness of unbound HVSs in the main \gaia catalogue as a function of mass. The different lines correspond to different limits on the heliocentric distance as indicated in the legend. Our observations are mostly sensitive to HVSs within 50 kpc.}
    \label{fig:mass_completeness_gaia}
\end{figure}

Using these results, we can also determine the completeness of \gaia for HVSs as a function of mass. In other words, what fraction of HVSs is expected to be in the \gaia catalogue as a function of mass. For this, we only consider HVSs that have not evolved into remnants. We show the results in Fig.~\ref{fig:mass_completeness_gaia} for varying distance limits.
We can see that the completeness for HVSs in \gaia approaches zero for masses below $\sim1 \; {\rm M_\odot}$ even when only considering HVSs within 50 kpc from the Sun, which are the ones our observations are mainly sensitive to. For high mass HVSs (M $\gtrsim 5$ M$_{\odot}$) the completeness converges to nearly one, because those stars do not live long enough to travel far from the GC.

\subsection{HVS parameter constraints}
\label{sec:HVS_parameter_constraints}
Having established the types of HVSs that \gaia is sensitive to, we will now describe our constraints on the ejection of HVSs.

By applying the same selections to the simulations described in Section~\ref{sec:model} as the observations, we can predict the number of HVSs that should be observed in our catalogue given a set of model parameters. By rejecting models that are incompatible with the observed number of HVSs, we provide constraints on the model parameters, and as such, the physical environment of the GC and the process of HVS ejection.

We use uniform priors within the parameter ranges given in Section~\ref{sec:simulations}. The posterior probability of the model with parameters $M$ is therefore given by
\begin{equation}
P(M | N_{\rm HVS}) \propto \frac{\lambda^{N_{\rm HVS}} \exp {-\lambda}}{N_{\rm HVS}!},  
\end{equation}
with $N_{\rm HVS}$ the number of observed HVSs and $\lambda$ the average number of expected HVSs according to the simulations for a given set of parameters $M$. 

We find that the main parameters we can constrain based on the number of observed HVSs are the ejection rate and MF \citep[as is also the case in][]{Evans_2022_I, Evans_2022_II}. Both the log-period distribution and mass-ratio distributions are significant in their effect on the observed number of HVSs, but affect the results much less than the MF and ejection rate. 

\begin{figure}
    \centering
    \includegraphics[width=\linewidth]{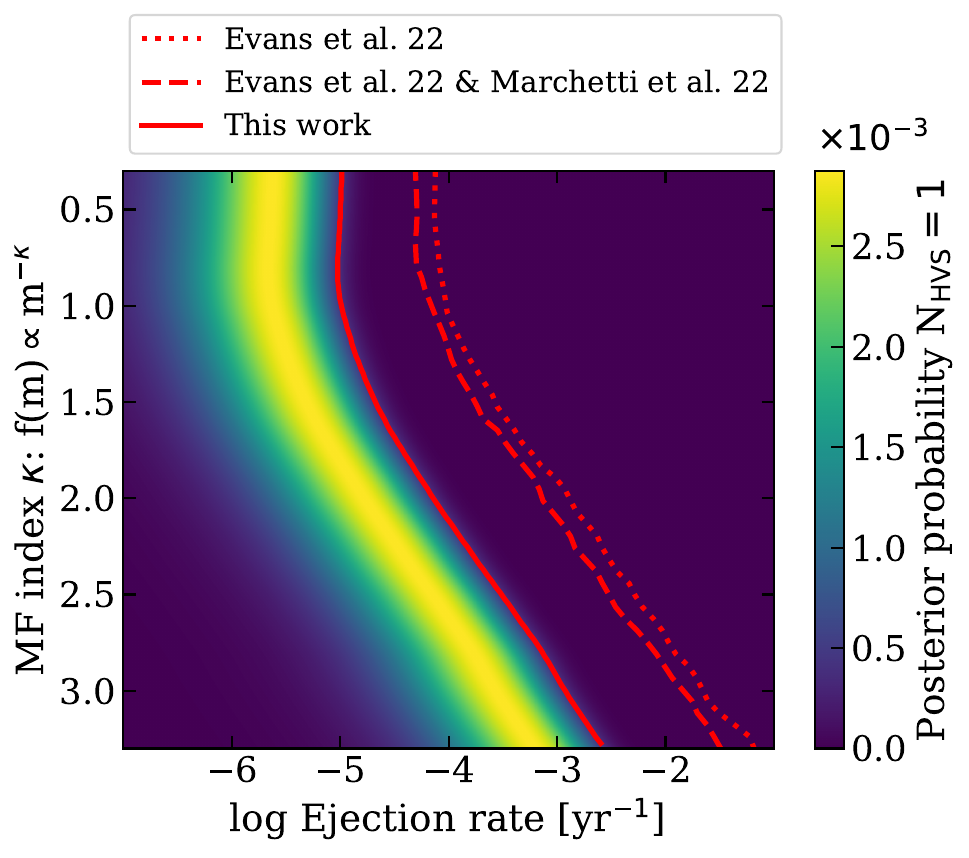}
    \caption{Posterior probability on the log of the HVS ejection rate and MF of the primary in the progenitor binary. The solid line gives the $2\sigma$ upper limit of the combined ejection rate and MF index. For comparison, the dotted and dashed lines give the $2\sigma$ upper limit on the same parameters from blind searches in the 6D phase-space \gaia eDR3 and DR3 samples respectively. The eDR3 constraint was calculated in \citet{Evans_2022_II} and the DR3 constraint is calculated by multiplying the posteriors in Fig.~3 of \citet{Marchetti_2022} with the top-right panel of Fig.~5 in \citet{Evans_2022_II}.}
    \label{fig:posterior_MF_ejection}
\end{figure}

In Fig.~\ref{fig:posterior_MF_ejection} we provide the posterior on the MF and ejection rate, marginalised over the binary period and mass ratio distributions.
Compared to previous studies, we see a remarkable improvement in the upper limit on the combined ejection rate and MF. Even though we only obtained 196 radial velocity measurements, our results are about one dex more constraining than the $\sim34$M radial velocity measurements in \gaia DR3 (also considering S5-HVS1 was discovered). This demonstrates the effectiveness of our targeted survey versus blind surveys when identifying HVSs. It is also worth noting that we use a more conservative prior on the log-period distribution compared to earlier work, since the prior used in \citet{Evans_2022_I, Evans_2022_II} results in higher ejection velocities to which our observations are most sensitive. We only highlight the upper limit and not the lower limit, because the upper limit is better constrained. The lower limit is highly sensitive to the prior used (e.g.~uniform or Gamma distribution). In Fig.~\ref{fig:app_posterior} we show the posterior if we use a log-uniform prior of $1/\lambda$. The posterior with the log-uniform prior shows that the upper limit remains, while allowing lower ejection rates with equal probability. 

For $\kappa\gtrsim 1.5$ we can see in Fig.~\ref{fig:posterior_MF_ejection} that the ejection rate is a strong function of the MF index. This is caused by our limited sensitivity to low-mass stars, as demonstrated in Fig.~\ref{fig:observable_by_gaia}. A more bottom-heavy MF would cause a lower fraction of HVSs to be observable by \gaia, thus necessitating a relatively higher ejection rate. For extremely top-heavy MFs, the ejection rate and MF index become relatively independent due to two competing effects: 
\begin{itemize}
    \item a larger fraction of HVSs is sufficiently massive to be observed with increasingly top-heavy MFs, and
    \item the finite lifetime of these increasingly massive stars causes a significant fraction of HVSs to evolve into stellar remnants before exiting the sphere within which they would be visible were they main sequence stars.
\end{itemize}

Overall, our upper limit ejection rate constraints are consistent with the process where stellar binaries' centre of mass trajectories have undergone gravitational scatterings off other (single) stars or giant molecular clouds: once every $10^{4}-10^{6}$ yr a binary star is sent on a highly eccentric orbit such that its pericenter becomes equal or smaller than the tidal separation radius (a.k.a. “loss cone orbit”) and an HVS is ejected \citep{Yu_2003, Perets_2007}. For a recent review on rates see \citet[][especially section 3.6.3]{Stone_2020}. Robust rate predictions to compare with our observational constraints should include realistic descriptions of the phase space and properties of binary stars, as well as those of single stars and giant molecular clouds out to $\sim 100$ pc from SgrA* in our GC. Such calculations and comparisons are beyond the scope of this paper and will be presented in a follow up investigation.

\begin{figure}
    \centering
    \includegraphics[width=\linewidth]{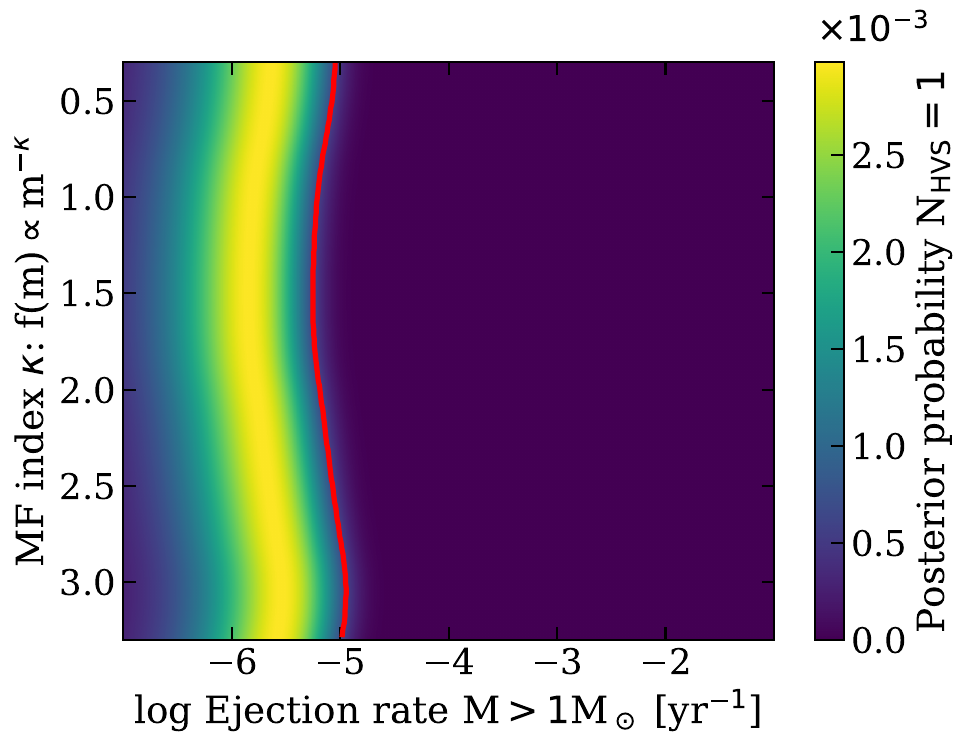}
    \caption{Same as Fig.~\ref{fig:posterior_MF_ejection}, but now only for HVSs with $M>1M_\odot$.}
    \label{fig:posterior_MF_ejection_observable}
\end{figure}

We can re-parameterise the ejection rate, such that it is no longer a function of the MF index by only considering HVSs massive enough to be observable by \gaia. The total ejection rate defines the ejection rate of HVSs in the mass range $[0.1, 100]\ \mathrm{M}_\odot$, as explained in Section~\ref{sec:simulations}. We can transform  this to the ejection rate in a different mass range by integrating the MF, which allows us to determine the number of stars in a mass interval $[M_1, M_2]$ 
\begin{equation}
\label{eq:mass_interval}
    N (M_1<M<M_2) \propto \frac{1}{\kappa+1} \left(M_2^{\kappa+1}-M_1^{\kappa+1}\right).
\end{equation}
The ejection rate for stars in the mass range $[M_1, M_2]$ is then the total ejection rate times the fraction of stars in the mass interval $[M_1, M_2]$. In Fig.~\ref{fig:posterior_MF_ejection_observable} we give the resulting posterior for HVSs more massive than $1\mathrm{M}_\odot$, making them potentially accessible to \gaia.
The figure demonstrates that irrespective of the MF, the ejection rate of HVSs more massive than $1\mathrm{M}_\odot$ cannot be higher than about $10^{-5}$ yr$^{-1}$ at $2\sigma$ confidence.

\begin{figure}
    \centering
    \includegraphics[width=\linewidth]{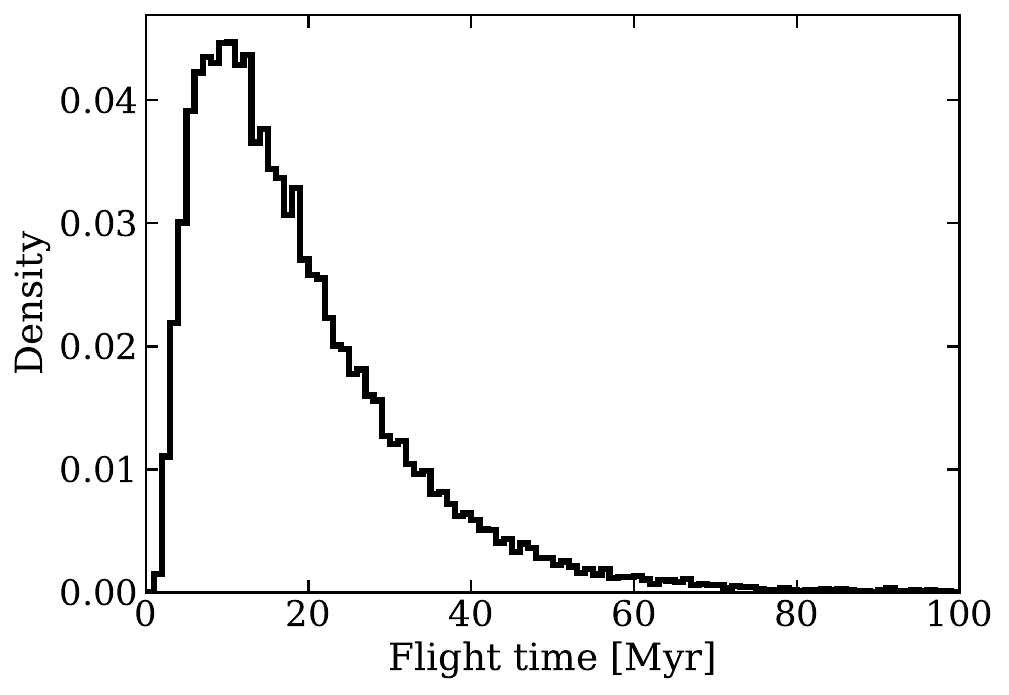}
    \caption{Predicted distribution of flight times for HVSs to which our observations are sensitive.}
    \label{fig:flight_times}
\end{figure}

In Fig.~\ref{fig:flight_times} we show the flight time (i.e.~time since disruption) distribution of HVSs that our observations are sensitive to.
We can see that the observations presented in this work are sensitive to ejections over the past $\sim 50-100$ Myr. Our ejection rate constraint therefore applies to ejections within the last $\sim 50-100$ Myr. HVS ejections that occurred more than 100 Myr ago would not produce any observable HVSs in our sample and can therefore not be constrained with our observations.

In the hypothetical case we observed two HVSs instead of one, the effect on our constraints is mainly that the posterior (Figs.~\ref{fig:posterior_MF_ejection} and \ref{fig:posterior_MF_ejection_observable}) becomes more narrow (the $2\sigma$ confidence interval by $\sim0.5$ dex). In addition, the $2\sigma$ upper limit constraint would increase by about $\sim0.1$ dex. 

\subsection{How many HVSs are in \gaia?}
\begin{figure}
    \centering
    \includegraphics[width=\linewidth]{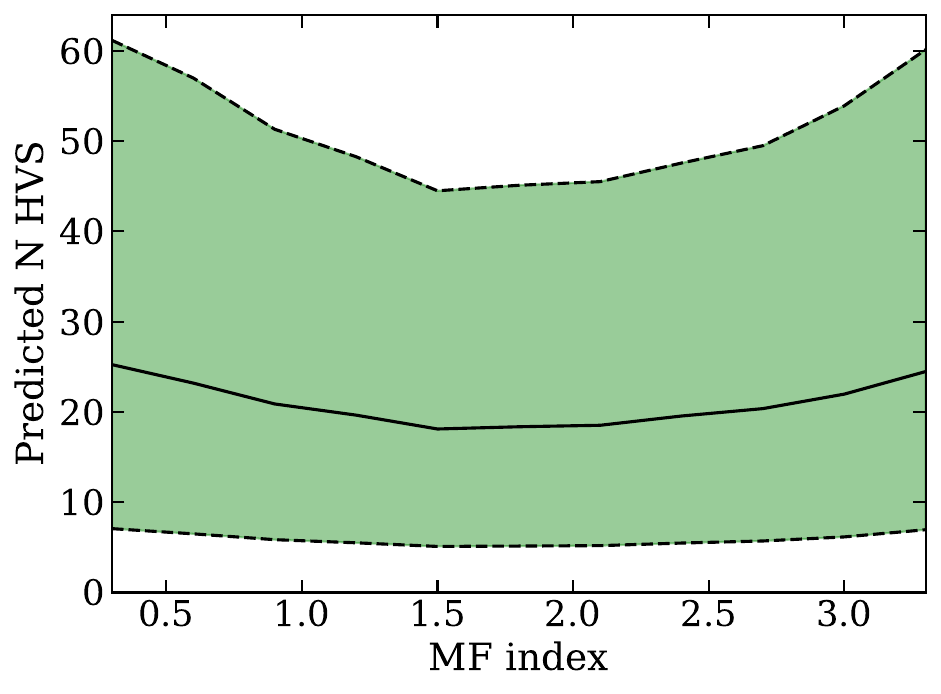}
    \caption{Number of unbound HVSs in the complete \gaia catalogue as a function of the MF of the progenitor binary population.}
    \label{fig:Gaia_N_HVS}
\end{figure}

Using our constraints from Section~\ref{sec:HVS_parameter_constraints}, we can predict how many HVSs are in the complete \gaia DR3 catalogue. Along with the predicted mass-distance relation, shown in Fig.~\ref{fig:observable_by_gaia}, this gives us a perspective on the discovery space of HVSs in \gaia. In Fig.~\ref{fig:Gaia_N_HVS} we show the expected number of HVSs in \gaia as a function of the MF.
Within $1\sigma$, we expect there to be between about 5 and 45 HVSs ($1.7 < \kappa < 2.3$), with the mode of the distribution around 18. The significant uncertainty in this number is caused by the single confirmed HVS within our observational sample. Importantly, the number of predicted HVSs is about one to two orders of magnitude less than predicted in earlier work \citep{Marchetti_2018}. This difference is mostly caused by a much lower ejection rate than previously expected. Theoretical prospects relying on \gaia discovered HVSs will have to take take this into account. 

\section{Discussion}
\label{sec:discussion}
Having presented our main results, we will now provide some discussion. We start by reviewing our assumptions in Section~\ref{sec:assumptions}, secondly we discuss the influence of the LMC on our results in Section~\ref{sec:LMC}, then we discuss the previously discovered HVS S5-HVS1 in light of our results in Section~\ref{sec:S5-HVS1}, we discuss some previously identified HVS candidates in Section~\ref{sec:distant_HVS_candidates}, and lastly provide prospects for the discovery of additional HVSs in Section~\ref{sec:prospects}.

\subsection{Assumptions}
\label{sec:assumptions}
In this work we rely on simulations to interpret our observational results. Evaluating the assumptions the simulations rely on is therefore important. Most of these have been discussed in \citet{Evans_2022_II}, including alternative MF and mass ratio parameterisations. We focus here on differences.

The MF we use in the simulations is different from the IMF for HVS progenitor binaries. The simulation assigns a random ejection time within the lifetime of a star, which makes the MF we constrain in this work more analogous to the present day mass function. In addition, the use of a single, time-independent MF in the simulations implicitly assumes a constant star formation rate. This is a simplifying assumption, which we know does not hold in detail \citep[e.g.][]{Nogueras_2020}.

Regarding the log-period distribution, we use a more stringent prior compared to \citet{Evans_2022_II}. Power-law indexes in $\alpha$ (see Section~\ref{sec:simulations}) below zero would result in higher ejection velocities because of the lower binary separation (see equation~\ref{eq:vej}). Since our observations are most sensitive to the fastest HVSs, our prior is more conservative.

We assumed a solar metallicity for the progenitor binaries throughout this work. We evaluate the effect of metallicity on our results by running identical simulations with $[{\rm Fe/H}]=0.3$. The difference on the posterior presented in Fig.~\ref{fig:posterior_MF_ejection} is about 7\% of the statistical uncertainty. We therefore conclude that metallicity is currently not an important consideration for our work.

We also tested the influence of the adopted Galactic potential on our results. To compare, we ran the simulations for the potential determined in \citet{McMillan_2017}. This potential is a more massive commonly used potential compared to our default one. The difference on our posterior from Fig.~\ref{fig:posterior_MF_ejection} changes by $0.01-0.1$ dex, which is well below the statistical uncertainty.

\subsection{Influence of the Large Magellanic Cloud}
\label{sec:LMC}
Our search for HVS candidates is based on pure radial trajectories. In reality, the non-spherical potential of the Galaxy deviates HVSs from exclusively radial trajectories. Because we compare our observations with simulations that are performed with a realistic, non-spherical potential and an identical selection procedure, these deviations from radial trajectories are not a concern for our results. However, the LMC is not incorporated in the potential model of the Galaxy in which the simulated HVSs are propagated. This might be a concern if the LMC significantly deviates HVSs from their trajectories \citep[see][]{Kenyon_2018, Boubert_2020}. In that case, simulations over-predict the number of HVSs we would be able to find and therefore bias the ejection rate constraint towards values which are too small.

We can evaluate the influence of the LMC by a comparison between integrating orbits of simulated HVSs with and without the LMC. We reran the simulations for $\kappa=2.3$ including the LMC with a mass of $1.5\times10^{11}$ M$_\odot$, which is near the total mass most studies find and close to half the mass of the Galaxy within 50 kpc \citep{Erkal_2019, Vasiliev_2020, Shipp_2021, Koposov_2023}. In the run with the LMC, we found an increase of about $2.8\%$ in the number of recovered HVSs in the simulation, which is about half the Poisson noise for those simulations. Our simulation with the LMC did not include the reflex motion of the Galaxy, but since this effect is of the same order as the deflection \citep{Boubert_2020}, we conclude that the LMC has no discernible influence on our results. 

\subsection{S5-HVS1}
\label{sec:S5-HVS1}
The serendipitous discovery of S5-HVS1 in the \sssss survey  \citep{li2019} appears curious. It is the only HVS that can be unambiguously traced back to the GC and is, by far, the fastest out of the stars that are suggested to originate from the Hills mechanism, in addition to being the closest \citep{Koposov_2020, Brown_2014, Brown_2018}. An interesting question to pose is then: what is the chance that the \sssss survey included an HVS? To understand this likelihood, we again make use of the simulations described in Section~\ref{sec:simulations}. 

We consider the footprint of the \sssss survey to be any stars with \gaia parallax $\varpi < 3\sigma_\varpi + 0.2$, mock DECam \citep{DECam_2016} photometry $15 < g < 19.5$, $-0.4 < (g-r) < 0.1$, and falling within 2 deg from an \sssss pointing \citep[see][table 2]{Li_2019} following \citet{Evans_2022_II}. Only HVSs with heliocentric radial velocities of $V_{\text{r}} > 800$ \kms are considered, since these were inspected in \citet{Koposov_2020}. We find that within the $1\sigma$ posterior probability on the MF and ejection rate from Fig.~\ref{fig:posterior_MF_ejection}, the number of expected HVSs in \sssss is about $0.14_{-0.11}^{+0.36}$. Although unlikely, the discovery of S5-HVS1 in \sssss is consistent with our constraints on the ejection rate of HVSs.

The radial velocity we measure for S5-HVS1 is $995\pm12$ \kms when considering the wavelength range 3850 to 5240 \AA. We used this wavelength range in particular because the spectrum we obtained for S5-HVS1 was much higher signal-to-noise than the other sources, and we noticed the wavelength calibration below 3850 \AA\ showed systematics. This is most likely caused by the lack of calibration lines being available for that part of the spectrum. Our measurement is consistent with the earlier found $1017\pm2.7$ \kms, so we find no indication of binarity.

Throughout this work, we assume that S5-HVS1 has been ejected by the Hills mechanism. Other mechanisms have, however, not been ruled out. A three-body interaction between a hypothetical intermediate mass black hole and Sgr A$^*$, for instance, could have ejected S5-HVS1 and is consistent with the S-star eccentricity distribution \citep{Generozov_2020}. However, \citet{Evans_2023} find that the lack of further detected HVSs in \gaia would necessitate a very specific configuration of any hypothetical intermediate mass black hole and Sgr A$^*$.

\subsection{Distant HVS candidates}
\label{sec:distant_HVS_candidates}
In general, the predicted HVSs in \gaia (see Fig.~\ref{fig:observable_by_gaia}) appear similar to those found in \citet{Brown_2014}, being dominated by distant stars of a few M$_\odot$. The difficulty with these stars is that due to their distance, their orbits can not be confidently traced back to the GC \citep{Irrgang_2018, Kreuzer_2020}. Confusion with other types of sources, such as disc runaways and halo stars is therefore an issue \citep[e.g.][]{Brown_2018}.

\begin{figure}
    \centering
    \includegraphics[width=\linewidth]{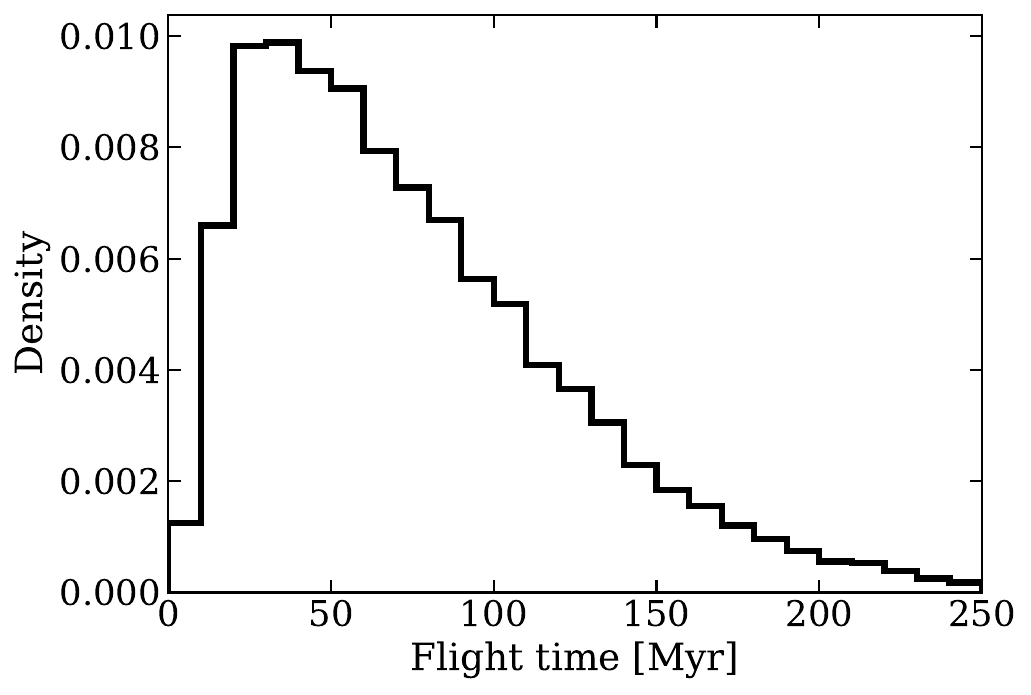}
    \caption{Flight time distribution for simulated HVSs matching the MMT survey selections.}
    \label{fig:MMT_flight_time}
\end{figure}

We use the posterior on the ejection rate from Fig.~\ref{fig:posterior_MF_ejection} to make a prediction on the number of HVSs within the MMT HVS Survey \citep{Brown_2014}. We reproduce the same colour selections outlined in section 2.1 of \citet{Brown_2012}. For simplicity, we consider the footprint of \textit{SDSS} to be ${\rm Dec}>0$ deg for the Northern Galactic cap, and ${\rm Dec} > -15$ deg for the Southern Galactic cap with a Galactic latitude $|l|>30$ deg \citep[for the sky coverage see][]{Aihara_2011}. \citet{Brown_2018} only studies stars with a heliocentric radial velocity transformed to the Galactic frame greater than $275$ \kms, which we follow. In their study, \citet{Brown_2018} identify seven stars with a probable origin in the GC. We use the posterior from Fig.~\ref{fig:posterior_MF_ejection}, in combination with the above selections to calculate how many HVSs there are predicted to be within this footprint. We sample the MF index in the range $[1.7, 2.3]$ and multiply each sample by the posterior on the ejection rate for that MF index. We find that within a 68\% confidence interval, we expect there to be between $0.3$ and $2.8$ HVSs in the MMT survey footprint. For the 95\% confidence interval, we expect there to be between $0.1$ and $5.1$ HVSs respectively. It is thus likely that a number of the stars identified in \citet{Brown_2018} are genuine HVSs. Interestingly however, there is thus a $>2\sigma$ tension between our results and the number of GC origin HVSs found in \citet{Brown_2018}. Apart from a statistical deviation, we have two main hypotheses as to the origin of this tension: the ejection rate of HVSs has not been constant over the flight time of the stars identified in \citet{Brown_2018}, or (a fraction of) the stars identified in \citet{Brown_2018} have a different origin \citep[supported by][]{Generozov_2022}. To investigate the first hypothesis, we show the distribution of flight times for predicted HVSs in the MMT footprint in Fig.~\ref{fig:MMT_flight_time} .
By comparison to Fig.~\ref{fig:flight_times}, we can clearly see that the simulated HVSs matching the MMT survey selections tend to have been ejected longer ago relative to those our survey presented in this work is sensitive to. The unbound HVS candidates reported to have a probable origin in the GC have flight times between about 50 and 150 Myr \citep[see][table 1]{Brown_2014}, which is consistent with our predictions.

\citet{Brown_2018} estimated the ejection rate of unbound HVSs in the mass range $2.5 \leq M_\odot \leq 4$ to be $1.5\times10^{-6}$ yr$^{-1}$. Using our observations, we can provide an independent constraint on the ejection rate in this mass range. We do this following the same procedure as for Fig.~\ref{fig:posterior_MF_ejection_observable}, but now with the above mentioned mass range. Our constraints on this ejection rate are relatively insensitive to the MF varying by a factor of about 3. The $2\sigma$ upper limit peaks at about $1.3\times10^{-6}$ yr$^{-1}$ for an MF index $\kappa\sim2.5$, while the mode of the posterior peaks at about $2.8\times10^{-7}$ yr$^{-1}$ (at the same MF index). Our maximum likelihood prediction is thus more than five times lower compared to the findings in \citet{Brown_2018}.

\subsection{Prospects}
\label{sec:prospects}
\gaia DR4 will include radial velocities for sources with a limiting magnitude of G$_{\mathrm{RVS}}<16.2$ \citep{Katz_2023}. If we consider all unbound HVSs with our ejection rate constraints from Section~\ref{sec:constraints}, we predict there to be about $3^{+4}_{-2}$ HVSs within the \gaia DR4 radial velocity catalogue if the MF index is between $1.7$ and $2.3$. It is, however, unlikely that all hypothetical HVSs in the footprint of the radial velocity catalogue of \gaia DR4 could be recognised as such. If we additionally require an accurate parallax measurement ($\varpi/\sigma_{\varpi}>5$), we only expect there to be $1^{+1}_{-1}$ HVSs. If on top of that, we only consider HVSs with radial velocities $>500$ \kms, we are left with fewer than one expected HVS.

A potentially promising means to extend to fainter stars is using radial velocities from the low-resolution \gaia\ ${\rm G_{BP}/G_{RP}}$ spectra \citep{Verberne_2024}. These spectra will be published for all \gaia sources in DR4. The radial velocity measurements from these spectra are most accurate for blue sources (${\rm G_{BP}-G_{RP} \lesssim0.7}$), which is the expected main discovery space for HVSs in \gaia (see Fig.~\ref{fig:observable_by_gaia}). Because the nominal uncertainties are high for radial velocity measurements using the ${\rm G_{BP}/G_{RP}}$ spectra, we only consider HVSs with radial velocities greater than 1000 \kms, in addition to only selecting stars with ${\rm G_{BP}-G_{RP}}<0.7$. Within this parameter range, we expect there to be $7^{+9}_{-5}$ HVSs in \gaia DR4. However, significant improvements would be required in the technique of obtaining these radial velocities from the ${\rm G_{BP}/G_{RP}}$ spectra to reliably identify these HVSs \citep{Verberne_2024}.\\

In addition to \gaia DR4, large upcoming spectroscopic survey instruments such as {\it DESI} \citep{Cooper_2023}, {\it WEAVE} \citep{Dalton_2014} and {\it 4MOST} \citep{deJong_2019} provide the opportunity to observe much larger numbers of sources. As part of the low-resolution high-latitude survey, {\it WEAVE} will observe about $20\,000$ HVS candidates in the northern hemisphere. This is two orders of magnitude larger than the observational sample described in this work. Moreover, {\it WEAVE} will allow us to observe stars down to the detection limit of \gaia. The precise survey strategy for {\it WEAVE} will be discussed in upcoming work.

\section{Conclusion}
\label{sec:conclusion}
In this work, we presented a novel selection method of HVS candidates. Utilising ground-based follow-up observations, we were able to reject all newly identified candidates. By combining this null-detection of new HVSs with sophisticated simulations, we were able to significantly improve the constraints on both the ejection rate of HVSs and the MF of the HVS progenitors. We show that the MF and ejection rate are degenerate because \gaia is only  sensitive to HVSs more massive than about $1\mathrm{M_\odot}$. We are therefore able to provide robust constraints on the ejection rate for HVSs more massive than $1\mathrm{M_\odot}$ independent on the MF. Over the past $\sim50-100$ Myr, the average ejection rate for these stars cannot have been higher than about $10^{-5}$ yr$^{-1}$ at $2\sigma$ significance. 

We use our constraints on the ejection rate and MF to evaluate previously identified HVS candidates. We find evidence that either the ejection rate of HVSs has varied significantly over the past $\sim150$ Myr, or these previously identified candidates include stars that do not originate in the GC.

Our constraints predict that within a 68\% confidence interval, there are between about 5 and 45 unbound HVSs in the complete \gaia catalogue. The majority of these stars will be main-sequence stars with a mass of a few ${\rm M_\odot}$ and be at Heliocentric distances between $10-100$ kpc.

%%%%%%%%%%%%%%%%%%%%%%%%%%%%%%%%%%%%%%%%%%%%%%%%%%
\section*{Data Availability}
The \gaia data underlying this article were accessed from \url{https://gea.esac.esa.int/archive/}. The HVS candidate catalogue used in this work and the results from our analysis of the calibrated spectra can be accessed from \url{https://zenodo.org/doi/10.5281/zenodo.12179452}.

\section*{Acknowledgements}
The authors would like to thank C.A.L.~Bailer Jones for kindly making the source code to his work available and \mbox{Warren R.~Brown} for useful comments and suggestions.

This work is based on service observations made with the \INT (programme ING.NL.22B.002) operated on the island of La Palma by the Isaac Newton Group of Telescopes in the Spanish Observatorio del Roque de los Muchachos of the Instituto de Astrofísica de Canarias.

Based on observations collected at the European Southern Observatory under ESO programme(s) 110.23SU and 111.24MP.

EMR acknowledges support from European Research Council (ERC) grant number: 101002511/project acronym: VEGA\_P. SK acknowledges support from the Science \& Technology Facilities Council (STFC) grant ST/Y001001/1. TM acknowledges a European Southern Observatory (ESO) fellowship.  FAE acknowledges support from the University
of Toronto Arts \& Science Postdoctoral Fellowship program
and the Dunlap Institute.

This work has made use of data from the European Space Agency (ESA) mission
{\it Gaia} (\url{https://www.cosmos.esa.int/gaia}), processed by the {\it Gaia}
Data Processing and Analysis Consortium (DPAC,
\url{https://www.cosmos.esa.int/web/gaia/dpac/consortium}). Funding for the DPAC
has been provided by national institutions, in particular the institutions
participating in the {\it Gaia} Multilateral Agreement.
This paper made use of the Whole Sky Database (wsdb) created and maintained by Sergey Koposov at the Institute of Astronomy, Cambridge with financial support from the Science \& Technology Facilities Council (STFC) and the European Research Council (ERC). This research or product makes use of public auxiliary data provided by ESA/Gaia/DPAC/CU5 and prepared by Carine Babusiaux.

For the purpose of open access, the author has applied a Creative Commons Attribution (CC BY) licence to any Author Accepted Manuscript version arising from this submission.\\

Software: \texttt{NumPy} \citep{Harris_2020}, \texttt{SciPy} \citep{2020SciPy-NMeth}, \texttt{Matplotlib} \citep{Hunter_2007}, \texttt{Astropy} \citep{astropy:2013, astropy:2018, astropy:2022}, \texttt{dustmaps} \citep{Green_dustmaps}, \texttt{healpy} \citep{Gorski_2005, Zonca_2019}, \texttt{sqlutilpy}\footnote{\url{https://doi.org/10.5281/zenodo.6867957}}. This research made use of ccdproc, an Astropy package for image reduction \citep{matt_craig_2022_6533213}.


\bibliographystyle{mnras}
\bibliography{references} % if your bibtex file is called example.bib

\begin{thebibliography}{}
\makeatletter
\relax
\def\mn@urlcharsother{\let\do\@makeother \do\$\do\&\do\#\do\^\do\_\do\%\do\~}
\def\mn@doi{\begingroup\mn@urlcharsother \@ifnextchar [ {\mn@doi@}
  {\mn@doi@[]}}
\def\mn@doi@[#1]#2{\def\@tempa{#1}\ifx\@tempa\@empty \href
  {http://dx.doi.org/#2} {doi:#2}\else \href {http://dx.doi.org/#2} {#1}\fi
  \endgroup}
\def\mn@eprint#1#2{\mn@eprint@#1:#2::\@nil}
\def\mn@eprint@arXiv#1{\href {http://arxiv.org/abs/#1} {{\tt arXiv:#1}}}
\def\mn@eprint@dblp#1{\href {http://dblp.uni-trier.de/rec/bibtex/#1.xml}
  {dblp:#1}}
\def\mn@eprint@#1:#2:#3:#4\@nil{\def\@tempa {#1}\def\@tempb {#2}\def\@tempc
  {#3}\ifx \@tempc \@empty \let \@tempc \@tempb \let \@tempb \@tempa \fi \ifx
  \@tempb \@empty \def\@tempb {arXiv}\fi \@ifundefined
  {mn@eprint@\@tempb}{\@tempb:\@tempc}{\expandafter \expandafter \csname
  mn@eprint@\@tempb\endcsname \expandafter{\@tempc}}}

\bibitem[\protect\citeauthoryear{{Abolfathi} et~al.,}{{Abolfathi}
  et~al.}{2018}]{Abolfathi_2018}
{Abolfathi} B.,  et~al., 2018, \mn@doi [\apjs] {10.3847/1538-4365/aa9e8a},
  \href {https://ui.adsabs.harvard.edu/abs/2018ApJS..235...42A} {235, 42}

\bibitem[\protect\citeauthoryear{{Aihara} et~al.,}{{Aihara}
  et~al.}{2011}]{Aihara_2011}
{Aihara} H.,  et~al., 2011, \mn@doi [\apjs] {10.1088/0067-0049/193/2/29}, \href
  {https://ui.adsabs.harvard.edu/abs/2011ApJS..193...29A} {193, 29}

\bibitem[\protect\citeauthoryear{{Asplund}, {Grevesse}, {Sauval}  \&
  {Scott}}{{Asplund} et~al.}{2009}]{Asplund_2009}
{Asplund} M.,  {Grevesse} N.,  {Sauval} A.~J.,   {Scott} P.,  2009, \mn@doi
  [\araa] {10.1146/annurev.astro.46.060407.145222}, \href
  {https://ui.adsabs.harvard.edu/abs/2009ARA&A..47..481A} {47, 481}

\bibitem[\protect\citeauthoryear{{Astropy Collaboration} et~al.,}{{Astropy
  Collaboration} et~al.}{2013}]{astropy:2013}
{Astropy Collaboration} et~al., 2013, \mn@doi [\aap]
  {10.1051/0004-6361/201322068}, \href
  {http://adsabs.harvard.edu/abs/2013A%26A...558A..33A} {558, A33}

\bibitem[\protect\citeauthoryear{{Astropy Collaboration} et~al.,}{{Astropy
  Collaboration} et~al.}{2018}]{astropy:2018}
{Astropy Collaboration} et~al., 2018, \mn@doi [\aj] {10.3847/1538-3881/aabc4f},
  \href {https://ui.adsabs.harvard.edu/abs/2018AJ....156..123A} {156, 123}

\bibitem[\protect\citeauthoryear{{Astropy Collaboration} et~al.,}{{Astropy
  Collaboration} et~al.}{2022}]{astropy:2022}
{Astropy Collaboration} et~al., 2022, \mn@doi [apj] {10.3847/1538-4357/ac7c74},
  \href {https://ui.adsabs.harvard.edu/abs/2022ApJ...935..167A} {935, 167}

\bibitem[\protect\citeauthoryear{{Bailer-Jones}}{{Bailer-Jones}}{2015}]{Bailer-Jones_2015}
{Bailer-Jones} C. A.~L.,  2015, \mn@doi [\pasp] {10.1086/683116}, \href
  {https://ui.adsabs.harvard.edu/abs/2015PASP..127..994B} {127, 994}

\bibitem[\protect\citeauthoryear{{Bailer-Jones}, {Rybizki}, {Fouesneau},
  {Demleitner}  \& {Andrae}}{{Bailer-Jones} et~al.}{2021}]{Bailer-Jones_2021}
{Bailer-Jones} C.~A.~L.,  {Rybizki} J.,  {Fouesneau} M.,  {Demleitner} M.,
  {Andrae} R.,  2021, \mn@doi [\aj] {10.3847/1538-3881/abd806}, \href
  {https://ui.adsabs.harvard.edu/abs/2021AJ....161..147B} {161, 147}

\bibitem[\protect\citeauthoryear{{Bartko} et~al.,}{{Bartko}
  et~al.}{2010}]{Bartko_2010}
{Bartko} H.,  et~al., 2010, \mn@doi [\apj] {10.1088/0004-637X/708/1/834}, \href
  {https://ui.adsabs.harvard.edu/abs/2010ApJ...708..834B} {708, 834}

\bibitem[\protect\citeauthoryear{{Boehle} et~al.,}{{Boehle}
  et~al.}{2016}]{Boehle_2016}
{Boehle} A.,  et~al., 2016, \mn@doi [\apj] {10.3847/0004-637X/830/1/17}, \href
  {https://ui.adsabs.harvard.edu/abs/2016ApJ...830...17B} {830, 17}

\bibitem[\protect\citeauthoryear{{Boubert}, {Erkal}  \& {Gualandris}}{{Boubert}
  et~al.}{2020}]{Boubert_2020}
{Boubert} D.,  {Erkal} D.,   {Gualandris} A.,  2020, \mn@doi [\mnras]
  {10.1093/mnras/staa2211}, \href
  {https://ui.adsabs.harvard.edu/abs/2020MNRAS.497.2930B} {497, 2930}

\bibitem[\protect\citeauthoryear{{Brown}}{{Brown}}{2015}]{Brown_2015}
{Brown} W.~R.,  2015, \mn@doi [\araa] {10.1146/annurev-astro-082214-122230},
  \href {https://ui.adsabs.harvard.edu/abs/2015ARA&A..53...15B} {53, 15}

\bibitem[\protect\citeauthoryear{{Brown}, {Geller}  \& {Kenyon}}{{Brown}
  et~al.}{2012}]{Brown_2012}
{Brown} W.~R.,  {Geller} M.~J.,   {Kenyon} S.~J.,  2012, \mn@doi [\apj]
  {10.1088/0004-637X/751/1/55}, \href
  {https://ui.adsabs.harvard.edu/abs/2012ApJ...751...55B} {751, 55}

\bibitem[\protect\citeauthoryear{{Brown}, {Geller}  \& {Kenyon}}{{Brown}
  et~al.}{2014}]{Brown_2014}
{Brown} W.~R.,  {Geller} M.~J.,   {Kenyon} S.~J.,  2014, \mn@doi [\apj]
  {10.1088/0004-637X/787/1/89}, \href
  {https://ui.adsabs.harvard.edu/abs/2014ApJ...787...89B} {787, 89}

\bibitem[\protect\citeauthoryear{{Brown}, {Lattanzi}, {Kenyon}  \&
  {Geller}}{{Brown} et~al.}{2018}]{Brown_2018}
{Brown} W.~R.,  {Lattanzi} M.~G.,  {Kenyon} S.~J.,   {Geller} M.~J.,  2018,
  \mn@doi [\apj] {10.3847/1538-4357/aadb8e}, \href
  {https://ui.adsabs.harvard.edu/abs/2018ApJ...866...39B} {866, 39}

\bibitem[\protect\citeauthoryear{{Cantat-Gaudin} et~al.,}{{Cantat-Gaudin}
  et~al.}{2023}]{Cantat-Gaudin_2023}
{Cantat-Gaudin} T.,  et~al., 2023, \mn@doi [\aap]
  {10.1051/0004-6361/202244784}, \href
  {https://ui.adsabs.harvard.edu/abs/2023A&A...669A..55C} {669, A55}

\bibitem[\protect\citeauthoryear{{Choi}, {Dotter}, {Conroy}, {Cantiello},
  {Paxton}  \& {Johnson}}{{Choi} et~al.}{2016}]{Choi_2016}
{Choi} J.,  {Dotter} A.,  {Conroy} C.,  {Cantiello} M.,  {Paxton} B.,
  {Johnson} B.~D.,  2016, \mn@doi [\apj] {10.3847/0004-637X/823/2/102}, \href
  {https://ui.adsabs.harvard.edu/abs/2016ApJ...823..102C} {823, 102}

\bibitem[\protect\citeauthoryear{{Contigiani}, {Rossi}  \&
  {Marchetti}}{{Contigiani} et~al.}{2019}]{Contigiani_2019}
{Contigiani} O.,  {Rossi} E.~M.,   {Marchetti} T.,  2019, \mn@doi [\mnras]
  {10.1093/mnras/stz1547}, \href
  {https://ui.adsabs.harvard.edu/abs/2019MNRAS.487.4025C} {487, 4025}

\bibitem[\protect\citeauthoryear{{Cooper} et~al.,}{{Cooper}
  et~al.}{2023}]{Cooper_2023}
{Cooper} A.~P.,  et~al., 2023, \mn@doi [\apj] {10.3847/1538-4357/acb3c0}, \href
  {https://ui.adsabs.harvard.edu/abs/2023ApJ...947...37C} {947, 37}

\bibitem[\protect\citeauthoryear{Craig et~al.,}{Craig
  et~al.}{2022}]{matt_craig_2022_6533213}
Craig M.,  et~al., 2022, {astropy/ccdproc: 2.3.1 -- fixes astropy 5.1
  compatibility}, \mn@doi{10.5281/zenodo.6533213}, \url
  {https://doi.org/10.5281/zenodo.6533213}

\bibitem[\protect\citeauthoryear{{Dalton} et~al.,}{{Dalton}
  et~al.}{2014}]{Dalton_2014}
{Dalton} G.,  et~al., 2014, in {Ramsay} S.~K.,  {McLean} I.~S.,   {Takami} H.,
  eds,  Society of Photo-Optical Instrumentation Engineers (SPIE) Conference
  Series Vol. 9147, Ground-based and Airborne Instrumentation for Astronomy V.
  p. 91470L (\mn@eprint {arXiv} {1412.0843}), \mn@doi{10.1117/12.2055132}

\bibitem[\protect\citeauthoryear{{Dark Energy Survey Collaboration}
  et~al.,}{{Dark Energy Survey Collaboration} et~al.}{2016}]{DECam_2016}
{Dark Energy Survey Collaboration} et~al., 2016, \mn@doi [\mnras]
  {10.1093/mnras/stw641}, \href
  {https://ui.adsabs.harvard.edu/abs/2016MNRAS.460.1270D} {460, 1270}

\bibitem[\protect\citeauthoryear{{Do}, {Kerzendorf}, {Winsor}, {St{\o}stad},
  {Morris}, {Lu}  \& {Ghez}}{{Do} et~al.}{2015}]{Do_2015}
{Do} T.,  {Kerzendorf} W.,  {Winsor} N.,  {St{\o}stad} M.,  {Morris} M.~R.,
  {Lu} J.~R.,   {Ghez} A.~M.,  2015, \mn@doi [\apj]
  {10.1088/0004-637X/809/2/143}, \href
  {https://ui.adsabs.harvard.edu/abs/2015ApJ...809..143D} {809, 143}

\bibitem[\protect\citeauthoryear{{Do} et~al.,}{{Do} et~al.}{2019}]{Do_2019}
{Do} T.,  et~al., 2019, \mn@doi [Science] {10.1126/science.aav8137}, \href
  {https://ui.adsabs.harvard.edu/abs/2019Sci...365..664D} {365, 664}

\bibitem[\protect\citeauthoryear{{Dotter}}{{Dotter}}{2016}]{Dotter_2016}
{Dotter} A.,  2016, \mn@doi [\apjs] {10.3847/0067-0049/222/1/8}, \href
  {https://ui.adsabs.harvard.edu/abs/2016ApJS..222....8D} {222, 8}

\bibitem[\protect\citeauthoryear{Drimmel \& Poggio}{Drimmel \&
  Poggio}{2018}]{Drimmel_2018}
Drimmel R.,  Poggio E.,  2018, \mn@doi [Research Notes of the AAS]
  {10.3847/2515-5172/aaef8b}, 2, 210

\bibitem[\protect\citeauthoryear{{Dunstall} et~al.,}{{Dunstall}
  et~al.}{2015}]{Dunstall_2015}
{Dunstall} P.~R.,  et~al., 2015, \mn@doi [A&A] {10.1051/0004-6361/201526192},
  580, A93

\bibitem[\protect\citeauthoryear{Eisenhauer et~al.,}{Eisenhauer
  et~al.}{2005}]{Eisenhauer_2005}
Eisenhauer F.,  et~al., 2005, \mn@doi [The Astrophysical Journal]
  {10.1086/430667}, 628, 246

\bibitem[\protect\citeauthoryear{{Eisenhauer} et~al.,}{{Eisenhauer}
  et~al.}{2011}]{Eisenhauer_2011}
{Eisenhauer} F.,  et~al., 2011, The Messenger, \href
  {https://ui.adsabs.harvard.edu/abs/2011Msngr.143...16E} {143, 16}

\bibitem[\protect\citeauthoryear{{Erkal} et~al.,}{{Erkal}
  et~al.}{2019}]{Erkal_2019}
{Erkal} D.,  et~al., 2019, \mn@doi [\mnras] {10.1093/mnras/stz1371}, \href
  {https://ui.adsabs.harvard.edu/abs/2019MNRAS.487.2685E} {487, 2685}

\bibitem[\protect\citeauthoryear{{Evans}, {Marchetti}  \& {Rossi}}{{Evans}
  et~al.}{2022a}]{Evans_2022_I}
{Evans} F.~A.,  {Marchetti} T.,   {Rossi} E.~M.,  2022a, \mn@doi [\mnras]
  {10.1093/mnras/stac495}, \href
  {https://ui.adsabs.harvard.edu/abs/2022MNRAS.512.2350E} {512, 2350}

\bibitem[\protect\citeauthoryear{{Evans}, {Marchetti}  \& {Rossi}}{{Evans}
  et~al.}{2022b}]{Evans_2022_II}
{Evans} F.~A.,  {Marchetti} T.,   {Rossi} E.~M.,  2022b, \mn@doi [\mnras]
  {10.1093/mnras/stac2865}, \href
  {https://ui.adsabs.harvard.edu/abs/2022MNRAS.517.3469E} {517, 3469}

\bibitem[\protect\citeauthoryear{{Evans}, {Rasskazov}, {Remmelzwaal},
  {Marchetti}, {Castro-Ginard}, {Rossi}  \& {Bovy}}{{Evans}
  et~al.}{2023}]{Evans_2023}
{Evans} F.~A.,  {Rasskazov} A.,  {Remmelzwaal} A.,  {Marchetti} T.,
  {Castro-Ginard} A.,  {Rossi} E.~M.,   {Bovy} J.,  2023, \mn@doi [\mnras]
  {10.1093/mnras/stad2273}, \href
  {https://ui.adsabs.harvard.edu/abs/2023MNRAS.525..561E} {525, 561}

\bibitem[\protect\citeauthoryear{Everall, Boubert, Koposov, Smith  \&
  Holl}{Everall et~al.}{2021}]{Everall_2021}
Everall A.,  Boubert D.,  Koposov S.~E.,  Smith L.,   Holl B.,  2021, \mn@doi
  [Monthly Notices of the Royal Astronomical Society] {10.1093/mnras/stab041},
  502, 1908

\bibitem[\protect\citeauthoryear{{Feldmeier-Krause}, {Kerzendorf}, {Neumayer},
  {Sch{\"o}del}, {Nogueras-Lara}, {Do}, {de Zeeuw}  \&
  {Kuntschner}}{{Feldmeier-Krause} et~al.}{2017}]{Feldmeier_2017}
{Feldmeier-Krause} A.,  {Kerzendorf} W.,  {Neumayer} N.,  {Sch{\"o}del} R.,
  {Nogueras-Lara} F.,  {Do} T.,  {de Zeeuw} P.~T.,   {Kuntschner} H.,  2017,
  \mn@doi [\mnras] {10.1093/mnras/stw2339}, \href
  {https://ui.adsabs.harvard.edu/abs/2017MNRAS.464..194F} {464, 194}

\bibitem[\protect\citeauthoryear{{Fitzpatrick}}{{Fitzpatrick}}{1999}]{Fitzpatrick_1999}
{Fitzpatrick} E.~L.,  1999, \mn@doi [\pasp] {10.1086/316293}, \href
  {https://ui.adsabs.harvard.edu/abs/1999PASP..111...63F} {111, 63}

\bibitem[\protect\citeauthoryear{{GRAVITY Collaboration} et~al.,}{{GRAVITY
  Collaboration} et~al.}{2018}]{Gravity_2018}
{GRAVITY Collaboration} et~al., 2018, \mn@doi [\aap]
  {10.1051/0004-6361/201833718}, \href
  {https://ui.adsabs.harvard.edu/abs/2018A&A...615L..15G} {615, L15}

\bibitem[\protect\citeauthoryear{{GRAVITY Collaboration} et~al.,}{{GRAVITY
  Collaboration} et~al.}{2019}]{Gravity_2019}
{GRAVITY Collaboration} et~al., 2019, \mn@doi [\aap]
  {10.1051/0004-6361/201935656}, \href
  {https://ui.adsabs.harvard.edu/abs/2019A&A...625L..10G} {625, L10}

\bibitem[\protect\citeauthoryear{{Gaia Collaboration} et~al.,}{{Gaia
  Collaboration} et~al.}{2016}]{Gaia_2016}
{Gaia Collaboration} et~al., 2016, \mn@doi [\aap]
  {10.1051/0004-6361/201629272}, \href
  {https://ui.adsabs.harvard.edu/abs/2016A&A...595A...1G} {595, A1}

\bibitem[\protect\citeauthoryear{{Gaia Collaboration} et~al.,}{{Gaia
  Collaboration} et~al.}{2023}]{Gaia_2023}
{Gaia Collaboration} et~al., 2023, \mn@doi [\aap]
  {10.1051/0004-6361/202243940}, \href
  {https://ui.adsabs.harvard.edu/abs/2023A&A...674A...1G} {674, A1}

\bibitem[\protect\citeauthoryear{{Generozov} \& {Madigan}}{{Generozov} \&
  {Madigan}}{2020}]{Generozov_2020}
{Generozov} A.,  {Madigan} A.-M.,  2020, \mn@doi [\apj]
  {10.3847/1538-4357/ab94bc}, \href
  {https://ui.adsabs.harvard.edu/abs/2020ApJ...896..137G} {896, 137}

\bibitem[\protect\citeauthoryear{{Generozov} \& {Perets}}{{Generozov} \&
  {Perets}}{2022}]{Generozov_2022}
{Generozov} A.,  {Perets} H.~B.,  2022, \mn@doi [\mnras]
  {10.1093/mnras/stac1108}, \href
  {https://ui.adsabs.harvard.edu/abs/2022MNRAS.513.4257G} {513, 4257}

\bibitem[\protect\citeauthoryear{{Genzel}, {Eisenhauer}  \&
  {Gillessen}}{{Genzel} et~al.}{2010}]{Genzel_2010}
{Genzel} R.,  {Eisenhauer} F.,   {Gillessen} S.,  2010, \mn@doi [Reviews of
  Modern Physics] {10.1103/RevModPhys.82.3121}, \href
  {https://ui.adsabs.harvard.edu/abs/2010RvMP...82.3121G} {82, 3121}

\bibitem[\protect\citeauthoryear{Ghez et~al.,}{Ghez et~al.}{2003}]{Ghez_2003}
Ghez A.~M.,  et~al., 2003, \mn@doi [The Astrophysical Journal]
  {10.1086/374804}, 586, L127

\bibitem[\protect\citeauthoryear{Ghez et~al.,}{Ghez et~al.}{2008}]{Ghez_2008}
Ghez A.~M.,  et~al., 2008, \mn@doi [The Astrophysical Journal]
  {10.1086/592738}, 689, 1044

\bibitem[\protect\citeauthoryear{{Gillessen}, {Eisenhauer}, {Trippe},
  {Alexander}, {Genzel}, {Martins}  \& {Ott}}{{Gillessen}
  et~al.}{2009}]{Gillessen_2009}
{Gillessen} S.,  {Eisenhauer} F.,  {Trippe} S.,  {Alexander} T.,  {Genzel} R.,
  {Martins} F.,   {Ott} T.,  2009, \mn@doi [\apj]
  {10.1088/0004-637X/692/2/1075}, \href
  {https://ui.adsabs.harvard.edu/abs/2009ApJ...692.1075G} {692, 1075}

\bibitem[\protect\citeauthoryear{{Gillessen} et~al.,}{{Gillessen}
  et~al.}{2017}]{Gillessen_2017}
{Gillessen} S.,  et~al., 2017, \mn@doi [\apj] {10.3847/1538-4357/aa5c41}, \href
  {https://ui.adsabs.harvard.edu/abs/2017ApJ...837...30G} {837, 30}

\bibitem[\protect\citeauthoryear{{G{\'o}rski}, {Hivon}, {Banday}, {Wandelt},
  {Hansen}, {Reinecke}  \& {Bartelmann}}{{G{\'o}rski}
  et~al.}{2005}]{Gorski_2005}
{G{\'o}rski} K.~M.,  {Hivon} E.,  {Banday} A.~J.,  {Wandelt} B.~D.,  {Hansen}
  F.~K.,  {Reinecke} M.,   {Bartelmann} M.,  2005, \mn@doi [\apj]
  {10.1086/427976}, \href {http://adsabs.harvard.edu/abs/2005ApJ...622..759G}
  {622, 759}

\bibitem[\protect\citeauthoryear{Gould \& Quillen}{Gould \&
  Quillen}{2003}]{Gould_2003}
Gould A.,  Quillen A.~C.,  2003, \mn@doi [The Astrophysical Journal]
  {10.1086/375840}, 592, 935

\bibitem[\protect\citeauthoryear{{Green}}{{Green}}{2018}]{Green_dustmaps}
{Green} G.,  2018, \mn@doi [The Journal of Open Source Software]
  {10.21105/joss.00695}, \href
  {https://ui.adsabs.harvard.edu/abs/2018JOSS....3..695G} {3, 695}

\bibitem[\protect\citeauthoryear{{Guy} et~al.,}{{Guy} et~al.}{2023}]{Guy_2023}
{Guy} J.,  et~al., 2023, \mn@doi [\aj] {10.3847/1538-3881/acb212}, \href
  {https://ui.adsabs.harvard.edu/abs/2023AJ....165..144G} {165, 144}

\bibitem[\protect\citeauthoryear{{Habibi} et~al.,}{{Habibi}
  et~al.}{2017}]{habibi_17}
{Habibi} M.,  et~al., 2017, \mn@doi [\apj] {10.3847/1538-4357/aa876f}, \href
  {https://ui.adsabs.harvard.edu/abs/2017ApJ...847..120H} {847, 120}

\bibitem[\protect\citeauthoryear{Harris et~al.,}{Harris
  et~al.}{2020}]{Harris_2020}
Harris C.~R.,  et~al., 2020, \mn@doi [Nature] {10.1038/s41586-020-2649-2}, 585,
  357

\bibitem[\protect\citeauthoryear{{Hills}}{{Hills}}{1988}]{Hills_1988}
{Hills} J.~G.,  1988, \mn@doi [\nat] {10.1038/331687a0}, \href
  {https://ui.adsabs.harvard.edu/abs/1988Natur.331..687H} {331, 687}

\bibitem[\protect\citeauthoryear{Hunter}{Hunter}{2007}]{Hunter_2007}
Hunter J.~D.,  2007, \mn@doi [Computing in Science \& Engineering]
  {10.1109/MCSE.2007.55}, 9, 90

\bibitem[\protect\citeauthoryear{Hurley, Pols  \& Tout}{Hurley
  et~al.}{2000}]{Hurley_2000}
Hurley J.~R.,  Pols O.~R.,   Tout C.~A.,  2000, \mn@doi [Monthly Notices of the
  Royal Astronomical Society] {10.1046/j.1365-8711.2000.03426.x}, 315, 543

\bibitem[\protect\citeauthoryear{{Husser}, {Wende-von Berg}, {Dreizler},
  {Homeier}, {Reiners}, {Barman}  \& {Hauschildt}}{{Husser}
  et~al.}{2013}]{Husser_2013}
{Husser} T.~O.,  {Wende-von Berg} S.,  {Dreizler} S.,  {Homeier} D.,  {Reiners}
  A.,  {Barman} T.,   {Hauschildt} P.~H.,  2013, \mn@doi [\aap]
  {10.1051/0004-6361/201219058}, \href
  {https://ui.adsabs.harvard.edu/abs/2013A&A...553A...6H} {553, A6}

\bibitem[\protect\citeauthoryear{{Irrgang}, {Kreuzer}  \& {Heber}}{{Irrgang}
  et~al.}{2018}]{Irrgang_2018}
{Irrgang} A.,  {Kreuzer} S.,   {Heber} U.,  2018, \mn@doi [\aap]
  {10.1051/0004-6361/201833874}, \href
  {https://ui.adsabs.harvard.edu/abs/2018A&A...620A..48I} {620, A48}

\bibitem[\protect\citeauthoryear{{Irrgang}, {Dimpel}, {Heber}  \&
  {Raddi}}{{Irrgang} et~al.}{2021}]{Irrgang_2021}
{Irrgang} A.,  {Dimpel} M.,  {Heber} U.,   {Raddi} R.,  2021, \mn@doi [\aap]
  {10.1051/0004-6361/202040178}, \href
  {https://ui.adsabs.harvard.edu/abs/2021A&A...646L...4I} {646, L4}

\bibitem[\protect\citeauthoryear{{Katz} et~al.,}{{Katz}
  et~al.}{2023}]{Katz_2023}
{Katz} D.,  et~al., 2023, \mn@doi [\aap] {10.1051/0004-6361/202244220}, \href
  {https://ui.adsabs.harvard.edu/abs/2023A&A...674A...5K} {674, A5}

\bibitem[\protect\citeauthoryear{{Kenyon}, {Bromley}, {Brown}  \&
  {Geller}}{{Kenyon} et~al.}{2018}]{Kenyon_2018}
{Kenyon} S.~J.,  {Bromley} B.~C.,  {Brown} W.~R.,   {Geller} M.~J.,  2018,
  \mn@doi [\apj] {10.3847/1538-4357/aada04}, \href
  {https://ui.adsabs.harvard.edu/abs/2018ApJ...864..130K} {864, 130}

\bibitem[\protect\citeauthoryear{{Klessen}, {Spaans}  \& {Jappsen}}{{Klessen}
  et~al.}{2007}]{Klessen_2007}
{Klessen} R.~S.,  {Spaans} M.,   {Jappsen} A.-K.,  2007, \mn@doi [\mnras]
  {10.1111/j.1745-3933.2006.00258.x}, \href
  {https://ui.adsabs.harvard.edu/abs/2007MNRAS.374L..29K} {374, L29}

\bibitem[\protect\citeauthoryear{Kobayashi, Hainick, Sari  \& Rossi}{Kobayashi
  et~al.}{2012}]{Kobayashi_2012}
Kobayashi S.,  Hainick Y.,  Sari R.,   Rossi E.~M.,  2012, \mn@doi [The
  Astrophysical Journal] {10.1088/0004-637X/748/2/105}, 748, 105

\bibitem[\protect\citeauthoryear{{Koposov}}{{Koposov}}{2019}]{Koposov_2019}
{Koposov} S.~E.,  2019, {RVSpecFit: Radial velocity and stellar atmospheric
  parameter fitting}, Astrophysics Source Code Library, record ascl:1907.013
  (\mn@eprint {ascl} {1907.013})

\bibitem[\protect\citeauthoryear{{Koposov} et~al.,}{{Koposov}
  et~al.}{2020}]{Koposov_2020}
{Koposov} S.~E.,  et~al., 2020, \mn@doi [\mnras] {10.1093/mnras/stz3081}, \href
  {https://ui.adsabs.harvard.edu/abs/2020MNRAS.491.2465K} {491, 2465}

\bibitem[\protect\citeauthoryear{{Koposov} et~al.,}{{Koposov}
  et~al.}{2023}]{Koposov_2023}
{Koposov} S.~E.,  et~al., 2023, \mn@doi [\mnras] {10.1093/mnras/stad551}, \href
  {https://ui.adsabs.harvard.edu/abs/2023MNRAS.521.4936K} {521, 4936}

\bibitem[\protect\citeauthoryear{{Kreuzer}, {Irrgang}  \& {Heber}}{{Kreuzer}
  et~al.}{2020}]{Kreuzer_2020}
{Kreuzer} S.,  {Irrgang} A.,   {Heber} U.,  2020, \mn@doi [\aap]
  {10.1051/0004-6361/202037747}, \href
  {https://ui.adsabs.harvard.edu/abs/2020A&A...637A..53K} {637, A53}

\bibitem[\protect\citeauthoryear{{Kroupa}}{{Kroupa}}{2001}]{Kroupa_2001}
{Kroupa} P.,  2001, \mn@doi [\mnras] {10.1046/j.1365-8711.2001.04022.x}, \href
  {https://ui.adsabs.harvard.edu/abs/2001MNRAS.322..231K} {322, 231}

\bibitem[\protect\citeauthoryear{{Li} et~al.,}{{Li} et~al.}{2019a}]{li2019}
{Li} T.~S.,  et~al., 2019a, \mn@doi [\mnras] {10.1093/mnras/stz2731}, \href
  {https://ui.adsabs.harvard.edu/abs/2019MNRAS.490.3508L} {490, 3508}

\bibitem[\protect\citeauthoryear{{Li} et~al.,}{{Li} et~al.}{2019b}]{Li_2019}
{Li} T.~S.,  et~al., 2019b, \mn@doi [\mnras] {10.1093/mnras/stz2731}, \href
  {https://ui.adsabs.harvard.edu/abs/2019MNRAS.490.3508L} {490, 3508}

\bibitem[\protect\citeauthoryear{{L{\"o}ckmann}, {Baumgardt}  \&
  {Kroupa}}{{L{\"o}ckmann} et~al.}{2010}]{Lockmann_2010}
{L{\"o}ckmann} U.,  {Baumgardt} H.,   {Kroupa} P.,  2010, \mn@doi [\mnras]
  {10.1111/j.1365-2966.2009.15906.x}, \href
  {https://ui.adsabs.harvard.edu/abs/2010MNRAS.402..519L} {402, 519}

\bibitem[\protect\citeauthoryear{{Lu}, {Do}, {Ghez}, {Morris}, {Yelda}  \&
  {Matthews}}{{Lu} et~al.}{2013}]{Lu_2013}
{Lu} J.~R.,  {Do} T.,  {Ghez} A.~M.,  {Morris} M.~R.,  {Yelda} S.,   {Matthews}
  K.,  2013, \mn@doi [\apj] {10.1088/0004-637X/764/2/155}, \href
  {https://ui.adsabs.harvard.edu/abs/2013ApJ...764..155L} {764, 155}

\bibitem[\protect\citeauthoryear{{Maness} et~al.,}{{Maness}
  et~al.}{2007}]{Maness_2007}
{Maness} H.,  et~al., 2007, \mn@doi [\apj] {10.1086/521669}, \href
  {https://ui.adsabs.harvard.edu/abs/2007ApJ...669.1024M} {669, 1024}

\bibitem[\protect\citeauthoryear{{Marchetti}, {Contigiani}, {Rossi}, {Albert},
  {Brown}  \& {Sesana}}{{Marchetti} et~al.}{2018}]{Marchetti_2018}
{Marchetti} T.,  {Contigiani} O.,  {Rossi} E.~M.,  {Albert} J.~G.,  {Brown}
  A.~G.~A.,   {Sesana} A.,  2018, \mn@doi [\mnras] {10.1093/mnras/sty579},
  \href {https://ui.adsabs.harvard.edu/abs/2018MNRAS.476.4697M} {476, 4697}

\bibitem[\protect\citeauthoryear{{Marchetti}, {Evans}  \& {Rossi}}{{Marchetti}
  et~al.}{2022}]{Marchetti_2022}
{Marchetti} T.,  {Evans} F.~A.,   {Rossi} E.~M.,  2022, \mn@doi [\mnras]
  {10.1093/mnras/stac1777}, \href
  {https://ui.adsabs.harvard.edu/abs/2022MNRAS.515..767M} {515, 767}

\bibitem[\protect\citeauthoryear{{McMillan}}{{McMillan}}{2017}]{McMillan_2017}
{McMillan} P.~J.,  2017, \mn@doi [\mnras] {10.1093/mnras/stw2759}, \href
  {https://ui.adsabs.harvard.edu/abs/2017MNRAS.465...76M} {465, 76}

\bibitem[\protect\citeauthoryear{Moe \& Stefano}{Moe \&
  Stefano}{2013}]{Moe_2013}
Moe M.,  Stefano R.~D.,  2013, \mn@doi [The Astrophysical Journal]
  {10.1088/0004-637X/778/2/95}, 778, 95

\bibitem[\protect\citeauthoryear{Moe \& Stefano}{Moe \&
  Stefano}{2015}]{Moe_2015}
Moe M.,  Stefano R.~D.,  2015, \mn@doi [The Astrophysical Journal]
  {10.1088/0004-637X/810/1/61}, 810, 61

\bibitem[\protect\citeauthoryear{{Nandakumar}, {Ryde}, {Schultheis},
  {Thorsbro}, {J{\"o}nsson}, {Barklem}, {Rich}  \& {Fragkoudi}}{{Nandakumar}
  et~al.}{2018}]{Nandakumar_2018}
{Nandakumar} G.,  {Ryde} N.,  {Schultheis} M.,  {Thorsbro} B.,  {J{\"o}nsson}
  H.,  {Barklem} P.~S.,  {Rich} R.~M.,   {Fragkoudi} F.,  2018, \mn@doi
  [\mnras] {10.1093/mnras/sty1255}, \href
  {https://ui.adsabs.harvard.edu/abs/2018MNRAS.478.4374N} {478, 4374}

\bibitem[\protect\citeauthoryear{{Nogueras-Lara} et~al.,}{{Nogueras-Lara}
  et~al.}{2020}]{Nogueras_2020}
{Nogueras-Lara} F.,  et~al., 2020, \mn@doi [Nature Astronomy]
  {10.1038/s41550-019-0967-9}, \href
  {https://ui.adsabs.harvard.edu/abs/2020NatAs...4..377N} {4, 377}

\bibitem[\protect\citeauthoryear{{Paumard} et~al.,}{{Paumard}
  et~al.}{2006}]{Paumard_2006}
{Paumard} T.,  et~al., 2006, \mn@doi [\apj] {10.1086/503273}, \href
  {https://ui.adsabs.harvard.edu/abs/2006ApJ...643.1011P} {643, 1011}

\bibitem[\protect\citeauthoryear{{Pelupessy}, {van Elteren, A.}, {de Vries,
  N.}, {McMillan, S. L. W.}, {Drost, N.}  \& {Portegies Zwart, S.
  F.}}{{Pelupessy} et~al.}{2013}]{Pelupessy_2013}
{Pelupessy} F.~I.,  {van Elteren, A.} {de Vries, N.} {McMillan, S. L. W.}
  {Drost, N.}  {Portegies Zwart, S. F.} 2013, \mn@doi [A&A]
  {10.1051/0004-6361/201321252}, 557, A84

\bibitem[\protect\citeauthoryear{{Penoyre}, {Belokurov}  \& {Evans}}{{Penoyre}
  et~al.}{2022}]{Penoyre_2022}
{Penoyre} Z.,  {Belokurov} V.,   {Evans} N.~W.,  2022, \mn@doi [\mnras]
  {10.1093/mnras/stac959}, \href
  {https://ui.adsabs.harvard.edu/abs/2022MNRAS.513.2437P} {513, 2437}

\bibitem[\protect\citeauthoryear{{Perets}, {Hopman}  \& {Alexander}}{{Perets}
  et~al.}{2007}]{Perets_2007}
{Perets} H.~B.,  {Hopman} C.,   {Alexander} T.,  2007, \mn@doi [\apj]
  {10.1086/510377}, \href
  {https://ui.adsabs.harvard.edu/abs/2007ApJ...656..709P} {656, 709}

\bibitem[\protect\citeauthoryear{{Portegies Zwart} \& {McMillan}}{{Portegies
  Zwart} \& {McMillan}}{2018}]{PORTEGIESZWART_2018}
{Portegies Zwart} S.,  {McMillan} S.,  2018, {Astrophysical Recipes; The art of
  AMUSE}.
IOP Publishing, \mn@doi{10.1088/978-0-7503-1320-9}

\bibitem[\protect\citeauthoryear{{Portegies Zwart} et~al.,}{{Portegies Zwart}
  et~al.}{2009}]{PORTEGIESZWART_2009}
{Portegies Zwart} S.,  et~al., 2009, \mn@doi [New Astronomy]
  {https://doi.org/10.1016/j.newast.2008.10.006}, 14, 369

\bibitem[\protect\citeauthoryear{{Portegies Zwart}, McMillan, {van Elteren},
  Pelupessy  \& {de Vries}}{{Portegies Zwart}
  et~al.}{2013}]{PORTEGIESZWART_2013}
{Portegies Zwart} S.~F.,  McMillan S.~L.,  {van Elteren} A.,  Pelupessy F.~I.,
   {de Vries} N.,  2013, \mn@doi [Computer Physics Communications]
  {https://doi.org/10.1016/j.cpc.2012.09.024}, 184, 456

\bibitem[\protect\citeauthoryear{{Przybilla}, {Fernanda Nieva}, {Heber}  \&
  {Butler}}{{Przybilla} et~al.}{2008}]{Przybilla_2008}
{Przybilla} N.,  {Fernanda Nieva} M.,  {Heber} U.,   {Butler} K.,  2008,
  \mn@doi [\apjl] {10.1086/592245}, \href
  {https://ui.adsabs.harvard.edu/abs/2008ApJ...684L.103P} {684, L103}

\bibitem[\protect\citeauthoryear{Rossi, Kobayashi  \& Sari}{Rossi
  et~al.}{2014}]{Rossi_2014}
Rossi E.~M.,  Kobayashi S.,   Sari R.,  2014, \mn@doi [The Astrophysical
  Journal] {10.1088/0004-637X/795/2/125}, 795, 125

\bibitem[\protect\citeauthoryear{{Sana} et~al.,}{{Sana}
  et~al.}{2012}]{Sana_2012}
{Sana} H.,  et~al., 2012, \mn@doi [Science] {10.1126/science.1223344}, 337, 444

\bibitem[\protect\citeauthoryear{{Sana} et~al.,}{{Sana}
  et~al.}{2013}]{Sana_2013}
{Sana} H.,  et~al., 2013, \mn@doi [A&A] {10.1051/0004-6361/201219621}, 550,
  A107

\bibitem[\protect\citeauthoryear{Sari, Kobayashi  \& Rossi}{Sari
  et~al.}{2009}]{Sari_2010}
Sari R.,  Kobayashi S.,   Rossi E.~M.,  2009, \mn@doi [The Astrophysical
  Journal] {10.1088/0004-637X/708/1/605}, 708, 605

\bibitem[\protect\citeauthoryear{Schlafly \& Finkbeiner}{Schlafly \&
  Finkbeiner}{2011}]{Schlafly_2011}
Schlafly E.~F.,  Finkbeiner D.~P.,  2011, \mn@doi [The Astrophysical Journal]
  {10.1088/0004-637X/737/2/103}, 737, 103

\bibitem[\protect\citeauthoryear{{Schlegel}, {Finkbeiner}  \&
  {Davis}}{{Schlegel} et~al.}{1998}]{Schlegel_1998}
{Schlegel} D.~J.,  {Finkbeiner} D.~P.,   {Davis} M.,  1998, \mn@doi [\apj]
  {10.1086/305772}, \href
  {https://ui.adsabs.harvard.edu/abs/1998ApJ...500..525S} {500, 525}

\bibitem[\protect\citeauthoryear{{Sch{\"o}del}, {Merritt}  \&
  {Eckart}}{{Sch{\"o}del} et~al.}{2009}]{Schodel_2009}
{Sch{\"o}del} R.,  {Merritt} D.,   {Eckart} A.,  2009, \mn@doi [\aap]
  {10.1051/0004-6361/200810922}, \href
  {https://ui.adsabs.harvard.edu/abs/2009A&A...502...91S} {502, 91}

\bibitem[\protect\citeauthoryear{{Schultheis}, {Rich}, {Origlia}, {Ryde},
  {Nandakumar}, {Thorsbro}  \& {Neumayer}}{{Schultheis}
  et~al.}{2019}]{Schultheis_2019}
{Schultheis} M.,  {Rich} R.~M.,  {Origlia} L.,  {Ryde} N.,  {Nandakumar} G.,
  {Thorsbro} B.,   {Neumayer} N.,  2019, \mn@doi [\aap]
  {10.1051/0004-6361/201935772}, \href
  {https://ui.adsabs.harvard.edu/abs/2019A&A...627A.152S} {627, A152}

\bibitem[\protect\citeauthoryear{{Shipp} et~al.,}{{Shipp}
  et~al.}{2021}]{Shipp_2021}
{Shipp} N.,  et~al., 2021, \mn@doi [\apj] {10.3847/1538-4357/ac2e93}, \href
  {https://ui.adsabs.harvard.edu/abs/2021ApJ...923..149S} {923, 149}

\bibitem[\protect\citeauthoryear{{Stone}, {Vasiliev}, {Kesden}, {Rossi},
  {Perets}  \& {Amaro-Seoane}}{{Stone} et~al.}{2020}]{Stone_2020}
{Stone} N.~C.,  {Vasiliev} E.,  {Kesden} M.,  {Rossi} E.~M.,  {Perets} H.~B.,
  {Amaro-Seoane} P.,  2020, \mn@doi [\ssr] {10.1007/s11214-020-00651-4}, \href
  {https://ui.adsabs.harvard.edu/abs/2020SSRv..216...35S} {216, 35}

\bibitem[\protect\citeauthoryear{{Tody}}{{Tody}}{1986}]{Doug_1986}
{Tody} D.,  1986, in {Crawford} D.~L.,  ed.,  Society of Photo-Optical
  Instrumentation Engineers (SPIE) Conference Series Vol. 627, Instrumentation
  in astronomy VI. p.~733, \mn@doi{10.1117/12.968154}

\bibitem[\protect\citeauthoryear{{Tody}}{{Tody}}{1993}]{Dough_1993}
{Tody} D.,  1993, in {Hanisch} R.~J.,  {Brissenden} R.~J.~V.,   {Barnes} J.,
  eds,  Astronomical Society of the Pacific Conference Series Vol. 52,
  Astronomical Data Analysis Software and Systems II. p.~173

\bibitem[\protect\citeauthoryear{Vasiliev, Belokurov  \& Erkal}{Vasiliev
  et~al.}{2020}]{Vasiliev_2020}
Vasiliev E.,  Belokurov V.,   Erkal D.,  2020, \mn@doi [Monthly Notices of the
  Royal Astronomical Society] {10.1093/mnras/staa3673}, 501, 2279

\bibitem[\protect\citeauthoryear{{Verberne}, {Koposov}, {Rossi}, {Marchetti},
  {Kuijken}  \& {Penoyre}}{{Verberne} et~al.}{2024}]{Verberne_2024}
{Verberne} S.,  {Koposov} S.~E.,  {Rossi} E.~M.,  {Marchetti} T.,  {Kuijken}
  K.,   {Penoyre} Z.,  2024, \mn@doi [A&A] {10.1051/0004-6361/202348406}, 684,
  A29

\bibitem[\protect\citeauthoryear{Virtanen et~al.,}{Virtanen
  et~al.}{2020}]{2020SciPy-NMeth}
Virtanen P.,  et~al., 2020, \mn@doi [Nature Methods]
  {10.1038/s41592-019-0686-2}, \href {https://rdcu.be/b08Wh} {17, 261}

\bibitem[\protect\citeauthoryear{{Yu} \& {Tremaine}}{{Yu} \&
  {Tremaine}}{2003}]{Yu_2003}
{Yu} Q.,  {Tremaine} S.,  2003, \mn@doi [\apj] {10.1086/379546}, \href
  {https://ui.adsabs.harvard.edu/abs/2003ApJ...599.1129Y} {599, 1129}

\bibitem[\protect\citeauthoryear{{Zhao}, {Zhao}, {Chu}, {Jing}  \&
  {Deng}}{{Zhao} et~al.}{2012}]{LAMOST_2012}
{Zhao} G.,  {Zhao} Y.-H.,  {Chu} Y.-Q.,  {Jing} Y.-P.,   {Deng} L.-C.,  2012,
  \mn@doi [Research in Astronomy and Astrophysics]
  {10.1088/1674-4527/12/7/002}, \href
  {https://ui.adsabs.harvard.edu/abs/2012RAA....12..723Z} {12, 723}

\bibitem[\protect\citeauthoryear{Zonca, Singer, Lenz, Reinecke, Rosset, Hivon
  \& Gorski}{Zonca et~al.}{2019}]{Zonca_2019}
Zonca A.,  Singer L.,  Lenz D.,  Reinecke M.,  Rosset C.,  Hivon E.,   Gorski
  K.,  2019, \mn@doi [Journal of Open Source Software] {10.21105/joss.01298},
  4, 1298

\bibitem[\protect\citeauthoryear{{de Jong} et~al.,}{{de Jong}
  et~al.}{2019}]{deJong_2019}
{de Jong} R.~S.,  et~al., 2019, \mn@doi [The Messenger]
  {10.18727/0722-6691/5117}, \href
  {https://ui.adsabs.harvard.edu/abs/2019Msngr.175....3D} {175, 3}

\bibitem[\protect\citeauthoryear{{von Fellenberg} et~al.,}{{von Fellenberg}
  et~al.}{2022}]{Fellenberg_2022}
{von Fellenberg} S.~D.,  et~al., 2022, \mn@doi [\apjl]
  {10.3847/2041-8213/ac68ef}, \href
  {https://ui.adsabs.harvard.edu/abs/2022ApJ...932L...6V} {932, L6}

\makeatother
\end{thebibliography}

%%%%%%%%%%%%%%%%%%%%%%%%%%%%%%%%%%%%%%%%%%%%%%%%%%

%%%%%%%%%%%%%%%%% APPENDICES %%%%%%%%%%%%%%%%%%%%%
\clearpage
\appendix

\section{HR density of HVS candidates}
\begin{figure}
    \centering
    \includegraphics[width=\linewidth]{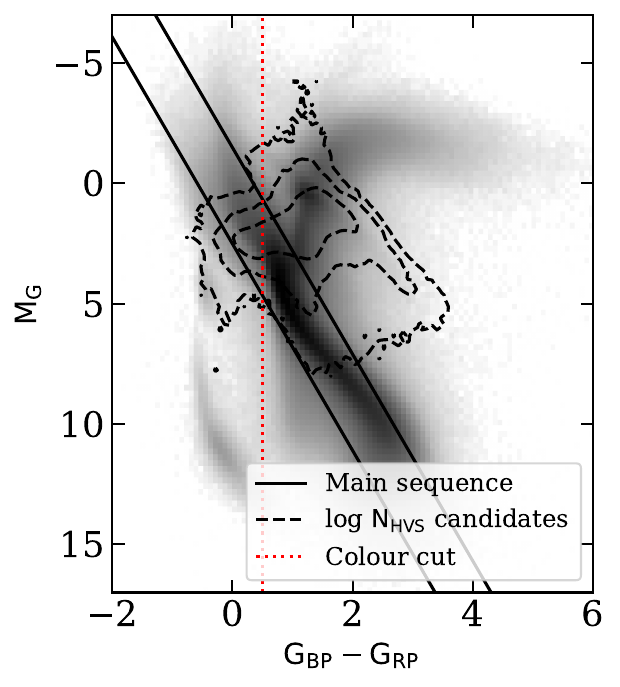}
    \caption{Same as Fig.~\ref{fig:HRD}, but now the overlain dashed line shows the density of HVS candidates that meet all our selections, except the colour-magnitude based ones.}
    \label{fig:app_HR}
\end{figure}

As mentioned in Section~\ref{sec:photo_selections}, the implied position for a star that is not moving on a radial trajectory can place it in a unphysical position in the HR diagram. In Fig.~\ref{fig:app_HR} we show where HVS candidates in \gaia that match all but our colour-magnitude selections end up in the HR diagram.

\section{Radial velocities from {\it SDSS} and {\it LAMOST}}
\label{app:external}
In Table~\ref{tab:external} we list the five sources for which we found literature radial velocity measurements in either {\it SDSS} or {\it LAMOST}.
\begin{table}
    \centering
    \caption{\gaia DR3 {\it source\_id}'s and radial velocities for the five stars for which we used literature radial velocity measurements.}
    \begin{tabular}{c|c|c}
         \hline
         \gaia DR3 {\it source\_id} & Radial velocity [\kms] & Source \\
         \hline
         2494761131458383872& $-185 \pm 13$ & {\it SDSS}\\
         2700705779669486848& $-73 \pm 10$& {\it LAMOST}\\
         3792400463887610368& $182 \pm 13$& {\it LAMOST}\\
         4430178054799074688& $197 \pm 8$& {\it SDSS}\\
         4459380675615776640& $-189 \pm 5$& {\it SDSS}\\
         \hline
    \end{tabular}
    \label{tab:external}
\end{table}

\section{G-magnitude distribution \gaia observable HVSs}
\begin{figure}
    \centering
    \includegraphics[width=\linewidth]{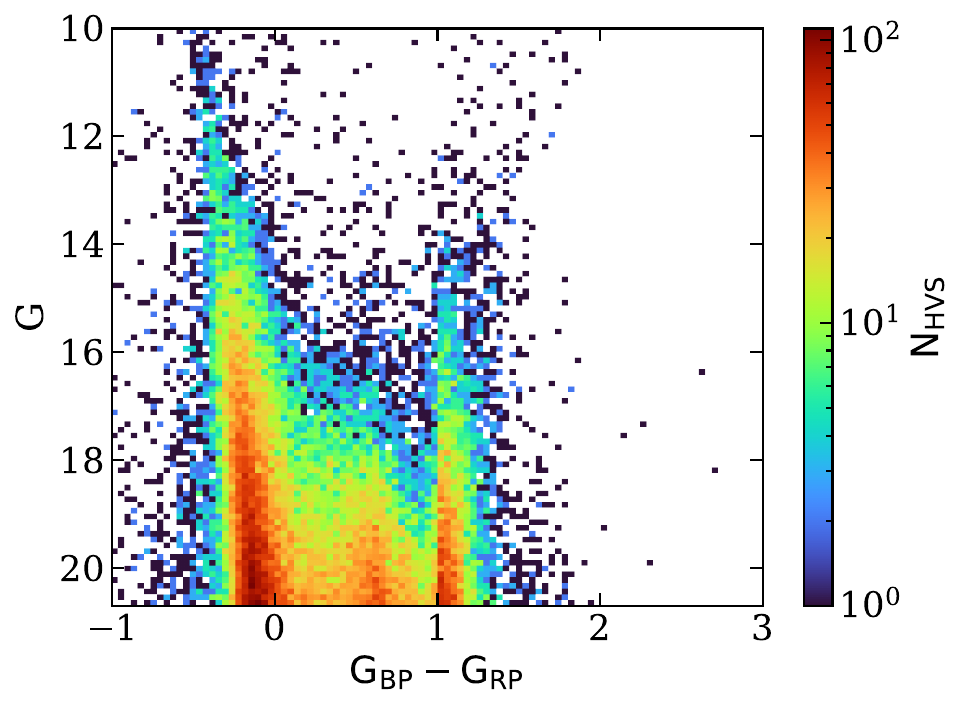}
    \caption{Apparent G-magnitude distribution of simulated HVSs predicted to be observable by \gaia.}
    \label{fig:app_G}
\end{figure}

In Fig.~\ref{fig:app_G} we show the apparent G-magnitude distribution for predicted HVSs that are observable by \gaia assuming an MF-index $\kappa=2.3$. The two vertical features correspond to the main-sequence and red-giant branch. We can see that most HVSs in \gaia will be faint, as expected, since the observed volume increases towards faint magnitudes.

\section{Lower limit on the ejection rate and MF}
\label{app:prior}
\begin{figure}
    \centering
    \includegraphics[width=\linewidth]{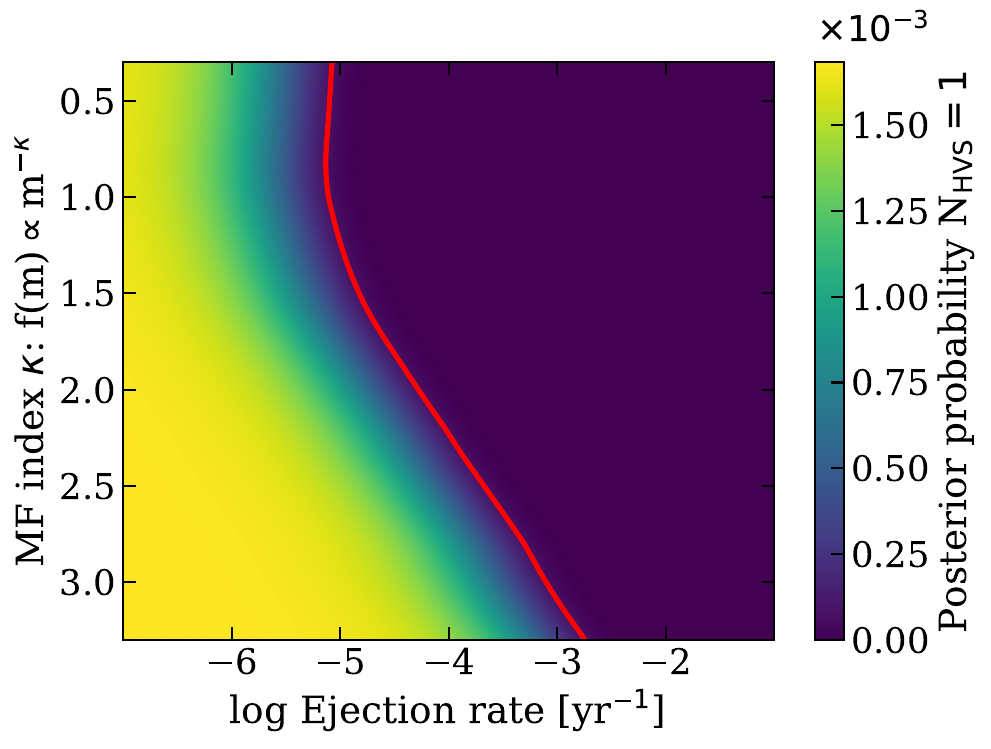}
    \caption{Posterior on the ejection rate and MF. Same as Fig.~\ref{fig:posterior_MF_ejection}, but with a log-uniform prior on the ejection rate of $1/\lambda$. The 95\% upper limit is indicated by the solid red line.}
    \label{fig:app_posterior}
\end{figure}

The lower limit on the ejection rate and MF is sensitive to the prior used. Throughout the main text, we used a uniform prior in $\lambda$. In Fig.~\ref{fig:app_posterior} we instead show the posterior if we use a log-uniform prior equal to $1/\lambda$. This prior might be more appropriate, since we do not know the magnitude of the rate, but do know it should be positive. This prior effectively removes the lower limit, since the posterior peaks for the limit of the ejection rate approaching zero.

%%%%%%%%%%%%%%%%%%%%%%%%%%%%%%%%%%%%%%%%%%%%%%%%%%

% Don't change these lines
\bsp	% typesetting comment
\label{lastpage}
\end{document}